\documentclass[journal,twoside,web]{ieeecolor}
\usepackage{generic}
\usepackage{cite}
\usepackage{amsmath,amssymb,amsfonts}
\usepackage{graphicx}
\usepackage{textcomp}
\usepackage{graphicx}
\usepackage{amsmath}
\usepackage{amssymb}
\usepackage{helvet}
\usepackage{courier}
\usepackage{float}
\usepackage{color}
\usepackage{booktabs}
\usepackage{multirow}
\usepackage{bbding}
\usepackage{pifont}
\usepackage{hyperref}
\usepackage[normalem]{ulem}
\usepackage{arydshln}
\usepackage{graphicx}
\usepackage{textcomp}
\usepackage{algorithmicx,algorithm}
\usepackage{multirow}
\usepackage{bbm}
\usepackage{graphicx}
\usepackage{amssymb}
\usepackage{helvet}
\usepackage{courier}
\usepackage{float}
\usepackage{color}
\usepackage{hyperref} 
\usepackage{amsmath,amsfonts}

\usepackage{epstopdf}

\usepackage{soul}
\soulregister\cite7 
\soulregister\ref7 

\usepackage{subcaption}
\usepackage{caption}
\def\BibTeX{{\rm B\kern-.05em{\sc i\kern-.025em b}\kern-.08em
    T\kern-.1667em\lower.7ex\hbox{E}\kern-.125emX}}
\pdfoutput=1

\begin{document}
\title{SkinFormer: Learning Statistical Texture Representation with Transformer for Skin Lesion Segmentation}
\author{Rongtao Xu,
        Changwei Wang,
        Jiguang Zhang,
        \\
        Shibiao Xu~\IEEEmembership{Member,~IEEE,}
        Weiliang Meng~\IEEEmembership{Member,~IEEE,}
        Xiaopeng Zhang,~\IEEEmembership{Member,~IEEE,}
\thanks{S. Xu and W. Meng are the corresponding authors (shibiaoxu@bupt.edu.cn; weiliang.meng@ia.ac.cn).}
\thanks{R. Xu is with the State Key Laboratory of Multimodal Artificial Intelligence Systems, Institute of Automation, Chinese Academy of Sciences, Beijing, China. C. Wang is with the Key Laboratory of Computing Power Network and Information Security, Ministry of Education, Shandong Computer Science Center, Qilu University of Technology, Jinan, China; Shandong Provincial Key Laboratory of Computer Networks, Shandong Fundamental Research Center for Computer Science, Jinan, China; CASIA, Beijing, China. J. Zhang, W. Meng, and X. Zhang are with the State Key Laboratory of Multimodal Artificial Intelligence Systems, Institute of Automation, Chinese Academy of Sciences. 
S. Xu is with school of Artificial Intelligence, Beijing University of Posts and Telecommunications.}
}

\maketitle

\begin{abstract}
Accurate skin lesion segmentation from dermoscopic images is of great importance for skin cancer diagnosis. However, automatic segmentation of melanoma remains a challenging task because it is difficult to incorporate useful texture representations into the learning process. Texture representations are not only related to the local structural information learned by CNN, but also include the global statistical texture information of the input image. In this paper, we propose a trans\textbf{Former} network (\textbf{SkinFormer}) that efficiently extracts and fuses statistical texture 
representation for \textbf{Skin} lesion segmentation. Specifically, to quantify the statistical texture of input features, a Kurtosis-guided Statistical Counting Operator is designed. We propose Statistical Texture Fusion Transformer and Statistical Texture Enhance Transformer with the help of Kurtosis-guided Statistical Counting Operator by utilizing the transformer's global attention mechanism. The former fuses structural texture information and statistical texture information, and the latter enhances the statistical texture of multi-scale features. {Extensive experiments on three publicly available skin lesion datasets validate that our SkinFormer outperforms other SOAT methods, and our method achieves 93.2\% Dice score on ISIC 2018. It can be easy to extend SkinFormer to segment 3D images in the future.} Our code is available at \href{https://github.com/Rongtao-Xu/SkinFormer}{{https://github.com/Rongtao-Xu/SkinFormer}}.
\end{abstract}



\section{Introduction}
Skin cancer is one of the most prevalent tumors that affect the elderly~\cite{liu2021fcp}.
If treated properly, survival rates for patients can reach over 95\% with early identification~\cite{xie2016melanoma}.
Currently, dermatologists conduct further analysis almost exclusively by manually delineating areas of the skin lesion. The manual process is frequently time-consuming and subject to operator bias. In recent decades, dermatologists have been able to improve the clinical diagnosis of melanoma thanks to the advent of computer-aided diagnosis (CAD) systems. {However, the computer-aided diagnosis system of melanoma also faces the key challenge of high segmentation accuracy.} There is an urgent need in clinical practice to automatically segment object regions with high accuracy from dermoscopic images~\cite{gessert2019skin_tbme}.

\begin{figure}[htbp]
\begin{center}
  \includegraphics[width=1\linewidth]{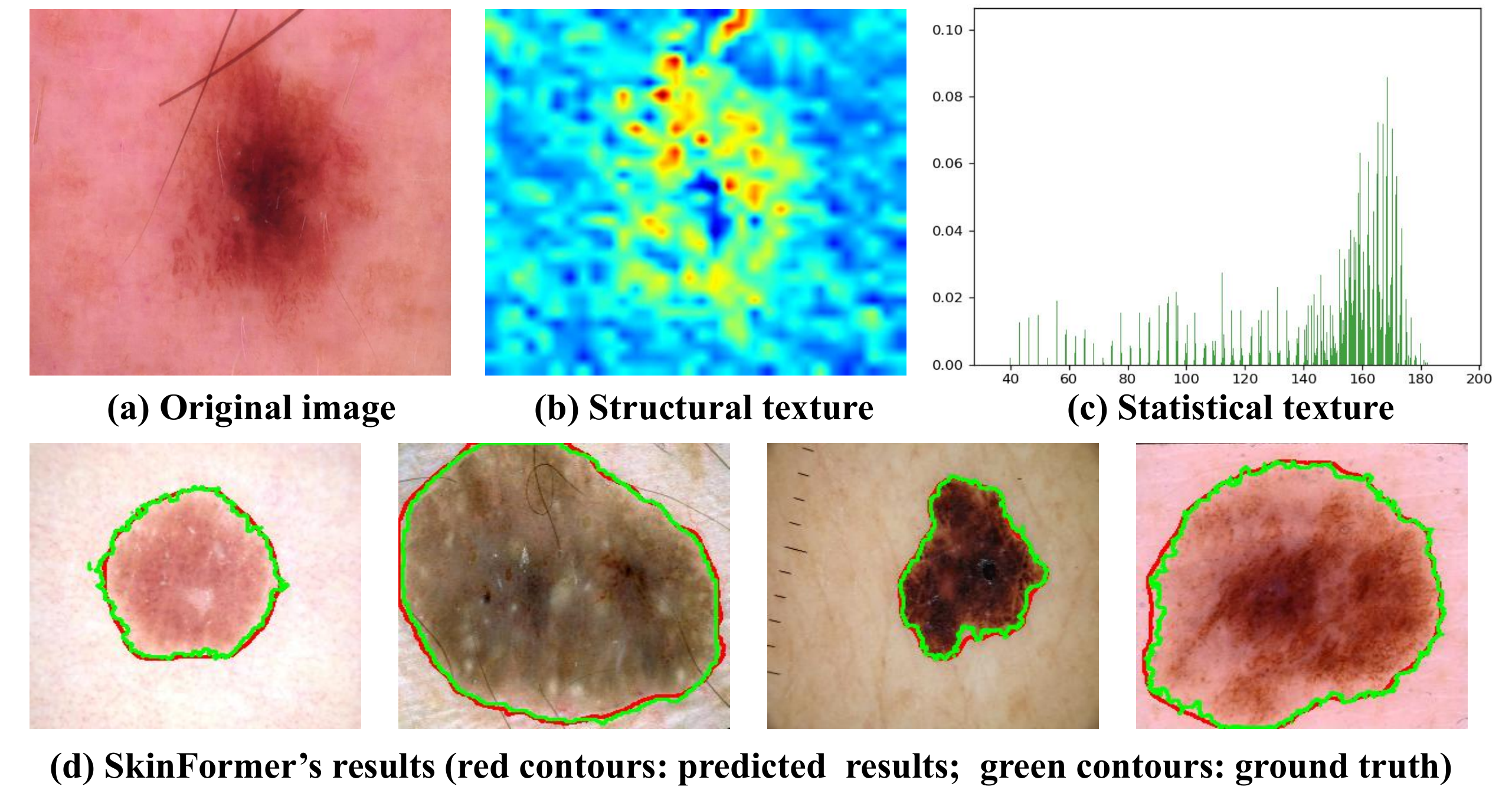}
\end{center}
   \caption{Examples of structural and statistical textures {and our results. (a) shows the original image. (b) shows the structure texture extracted by the typical convolutional neural network}~\cite{long2015fully}. (c) displays the histogram (statistical texture information) {of the original image. (d) Our SkinFormer's results. The red contours show the predicted results and the green contours show the ground truth.}}
\label{fig:introduoction}
\end{figure}

The segmentation task of the skin lesion is very challenging for four reasons: (1) The contrast between skin tissues is low, resulting in ambiguous boundaries of the lesion. (2) There are significant differences in the size, shape and color of skin lesions. (3) The same types of skin lesions are visually similar, but different types of skin lesions are visually different. (4) Dermoscopy images may contain interference factors such as hair and blood vessels. To address these challenges, many algorithms based on deep convolutional networks ~\cite{bi2019step} have made significant progress.

There is a complex correlation between skin lesions and their surrounding contextual regions as skin lesions gradually invade adjacent tissues. Therefore it is crucial to incorporate useful texture representations into the learning process. Current skin lesion segmentation methods mainly extract context information in high-level features. High-level features in deep layers often lead to inaccurate outputs due to the loss of texture information in low-level features. It is currently popular to use skip connections to fuse low-level and high-level features. U-Net~\cite{Unet} uses the connection of low-level and high-level features with different scales to improve the accuracy of medical image segmentation tasks. DeepLabv3+~\cite{chen2018DeepLabV3} directly fuses shallow and deep features as the input for predicting heads. These methods validate that structural texture information in the shallow layers of CNNs is crucial for segmentation tasks, especially on medical images with blurred edge details.

Image textures contain not only local structural properties but also global statistical properties~\cite{haralick1973textural,castleman1996digital}. As shown in Figure~\ref{fig:introduoction}, in addition to the structural texture extracted by CNN, another important property of texture is the statistical texture, such as the distribution histogram of the analyzed image. Many methods often only focus on the structural texture information in the shallow layers of CNN, and do not effectively use and fuse statistical texture information in medical images for semantic segmentation. In contrast, we propose a Statistical Texture Transformer Network for skin lesion segmentation, named `SkinFormer'. 
{First, we present a Kurtosis-guided Statistical Counting Operator to describe statistical texture information. Specifically, since convolution operations in deep neural networks are difficult to extract and optimize the statistical texture of images, we propose Kurtosis-guided Statistical Counting Operator to quantize input features into multiple levels.} Each level can represent a kind of statistical texture information. The feature map's kurtosis is then calculated because it reflects the sharpness. {And the kurtosis value of 0 indicates that it completely obeys the normal distribution.} We consider images with large absolute values of kurtosis (deviation from normal distribution) to have complex contextual information. We use kurtosis as a weight to make the network pay attention to the statistical texture information of images with large kurtosis values. After quantization, the intensity of each level is calculated.

With the help of the Kurtosis-guided Statistical Counting Operator, we design the Statistical Texture Fusion Transformer to effectively fuse the structural texture information and statistical texture information of medical images. {The Statistical Texture Fusion Transformer uses the comprehensive attention and local-window self-attention~\cite{yuan2021hrformer} to extract structural texture information}, and then controls the fusion degree of structural texture information and statistical texture information through a gating mechanism. 
{Comprehensive attention improves the representation ability of features by simultaneously paying attention to the information in the height and width directions of the features.}
{To increase the skin lesion segmentation's precision}, we further design a Statistical Texture Enhance Transformer, which can use the transformer to enhance multi-scale statistical texture information.
{The Statistical Texture Enhance Transformer takes the feature maps of multiple scales as input, and then employs the Multi-scale embedding enhancement and Texture-enhanced FFN to extract multi-scale statistical texture information.}

The contributions of this paper can be summarized as:
\begin{itemize}

\item {We propose a Kurtosis-guided Statistical Counting Operator to efficiently extract statistical texture in skin lesion images, which quantifies continuous input features into multiple levels and utilizes kurtosis to guide the network to focus on images with complex context.}

\item {We design the Statistical Texture Fusion Transformer, which uses a gating mechanism to control the degree of fusion for effectively fusing the structural texture information and statistical texture information. In addition, comprehensive attention is proposed to improve the representation ability of features.}

\item {We further design the Statistical Texture Enhance Transformer, which employs the Multi-scale embedding enhancement and Texture-enhanced FFN to enhance the extraction ability of multi-scale statistical textures representations.}

\item We propose a transformer network (SkinFormer) based on our Kurtosis-guided Statistical Counting Operator, Statistical Texture Fusion Transformer, and Statistical Texture Enhance Transformer, which can fully extract and fuse statistical texture information for skin lesion segmentation.

\end{itemize}

\section{Related work}
\subsection{Skin Lesion Segmentation}
Many semantic segmentation algorithms are applied to skin lesion segmentation, and these algorithms can be broadly classified into unsupervised and supervised methods~\cite{bi2017dermoscopic_tbme,glaister2014segmentation_tbme,sadri2012segmentation_tbme}. We focus on supervised methods~\cite{tartaglione2022loss}.
In the past decades, the challenge of skin lesion segmentation has been an important research topic~\cite{zhou2022h}.
Fully convolutional network~\cite{long2015fully} (FCN) based methods~\cite{yuan2017automatic,mirikharaji2018star} have advanced significantly in the task of skin lesion segmentation in recent years thanks to the development of deep learning.
Yuan et al.~\cite{yuan2017automatic} used a deep fully convolutional network with Jaccard distance for the skin lesion segmentation.
Mirikharaji et al.~\cite{mirikharaji2018star} gave a new loss term to train fully convolutional networks end-to-end.
Methods based on U-Net~\cite{Unet} and residual networks are also widely used in the field of skin lesion segmentation.
Zhou et al.~\cite{zhou2018unet++} proposed U-Net++ to combine the attention mechanism with U-Net.
Tang et al.~\cite{tang2019efficient} presented an image-based separable U-Net network with stochastic weights averaging.
Gu et al.~\cite{CE-Net} employed a general medical image segmentation network named CE-Net to extract image context information.
Tu et al.~\cite{qamar2021dense} combined the strengths of DenseNet and ResNet to improve the performance of skin lesion segmentation.
{In addition to these methods, it is essential to mention recent contributions such as FAT-Net and MEW-UNet in skin lesion segmentation. FAT-Net~\cite{wu2022fat} leverages attention mechanisms and feature adaptive transformers to achieve robust and accurate segmentation results. MEW-UNet~\cite{ruan2022mew} incorporates multi-axis representation learning into the U-Net architecture, enhancing its ability to capture detailed lesion features across different scales.}
Different from the previously mentioned medical image segmentation methods, our SkinFormer effectively extract and fuses statistical texture information, resulting in more accurate skin lesion segmentation.

\begin{figure*}[htbp]
\begin{center}
  \includegraphics[width=1\linewidth]{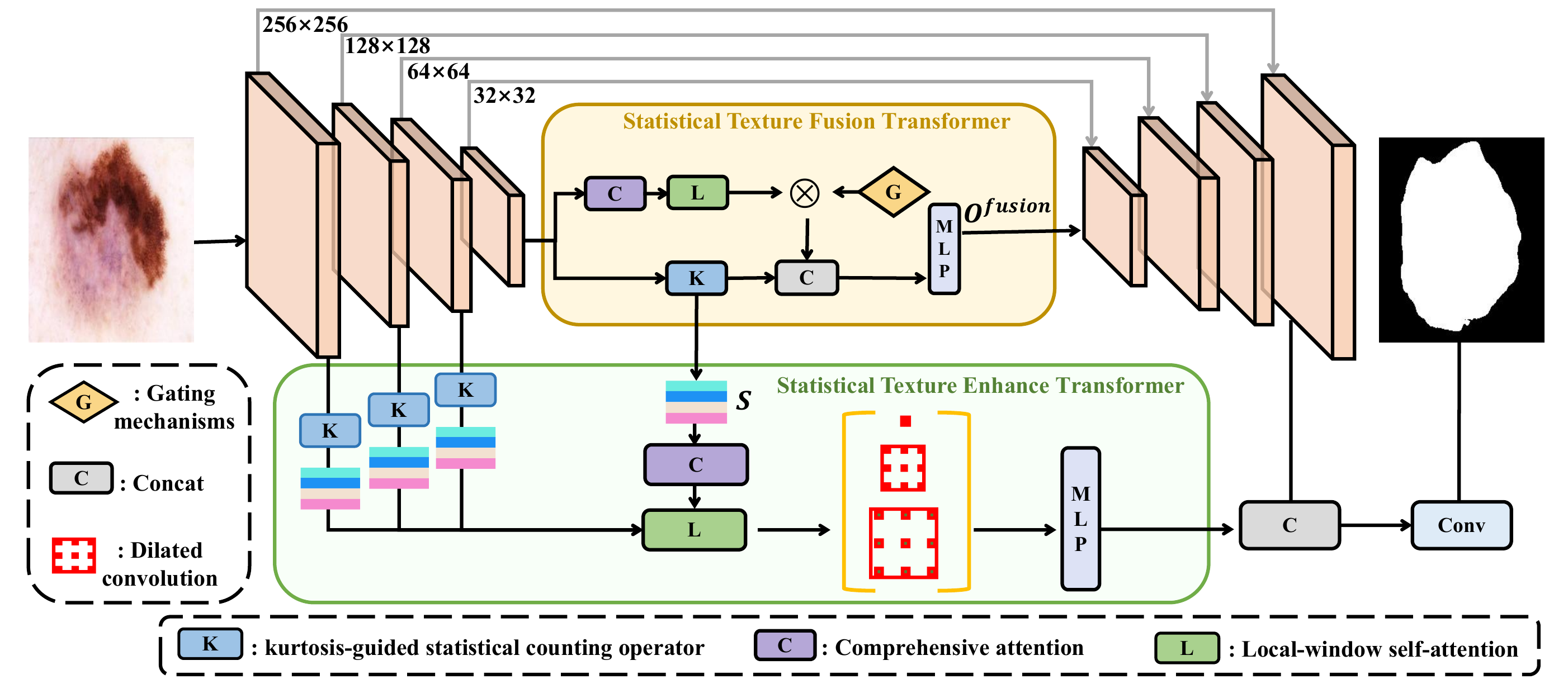}
\end{center}
   \caption{Overview of the our statistical texture transformer network (SkinFormer) with Kurtosis-guided Statistical Counting Operator, Statistical Texture Fusion Transformer and Statistical Texture Enhance Transformer. We adopt a U-shaped network structure to extract multi-level features. Then, the high-level features are fed into the Statistical Texture Fusion Transformer to obtain the statistical texture fusion feature $O^{fusion}$ and quantized intensity embedding $S$. {The multi-scale low-level features and $S$ are fed into the Statistical Texture Enhance Transformer to enhance the ability of texture representation extraction. The above two transformer-based modules both use comprehensive attention to improve the representation ability of features.}}
\label{fig:net}
\end{figure*}

\subsection{Transformer for Medical Segmentation}
Transformer was first successfully applied to natural language processing tasks.
ViT~\cite{dosovitskiy2020image} applies the transformer architecture to image classification tasks by serializing images into image patches, which inspires many image segmentation methods~\cite{xiao2023baseg}.
Transformer-based architectures~\cite{chen2022utrad} exploit a self-attention mechanism to encode long-range dependencies and have achieved excellent performance in medical image semantic segmentation tasks~\cite{xu2021dc,wang2022net}.
Valanarasu et al.~\cite{valanarasu2021medical} gave a method for medical image segmentation based on gated axial attention and transformers.
Chen et al.~\cite{chen2021transunet} proposed TransUNet, which combined U-Net architecture and transformer to achieve excellent performance on multi-organ segmentation and cardiac segmentation.
Xu et al.~\cite{xu2021dc} provided  DC-Net, which uses transformers to capture contextual information for medical image segmentation.
Hybrid architectures based on CNNs and transformers have been successful in the field of skin lesion segmentation, while transformer-based frameworks are difficult to achieve the same success because skin lesion segmentation usually has only thousands of data~\cite{wang2021boundary}.
Therefore, we also adopt the hybrid structure of CNN and transformer, and choose U-Net as our backbone network. Previous methods rarely pay attention to statistical texture information, and we are the first method to use transformers to fully extract and fuse statistical texture information according to our knowledge.

\subsection{Statistical Texture Representation Encoding}
The Kurtosis-guided Statistical Counting Operator is a technique for feature encoding. Common feature encoding methods are mainly aimed at the structural context information of images. Zhang et al.~\cite{zhang2018context} introduced a context encoding module to explore the impact of contextual information in semantic segmentation. Zhang et al.~\cite{zhang2017deep} proposed a deep texture encoding network for material and texture recognition. For the statistical texture information of images, many methods have validated that the feature encoding of statistical texture information can promote image understanding and recognition~\cite{wang2016learnable,xie2015effective,zhu2021learning}. Wang et al.~\cite{wang2016learnable} encoded features as learnable histograms, which achieve the goal of learning histogram features in deep neural networks in end-to-end training. Xie et al.~\cite{xie2015effective} gave a fast two-step texton encoding method to encode texture representations, and then fused two types of histogram features for classification.
{Multi-Scale Self-Guided Attention~\cite{sinha2020multi} and MALUNet~\cite{ruan2022malunet} are based on the idea of multi-scale feature extraction and fusion.
Different from these previous methods that rely solely on multi-scale feature extraction and fusion, our approach introduces a novel Kurtosis-guided Statistical Counting Operator to extract statistical texture information. Furthermore, our method includes the Statistical Texture Fusion Transformer, which effectively fuses structural and statistical texture information.}

\section{Method}
\label{Method}
In this section, we introduce our statistical texture transformer network (SkinFormer) for skin lesion segmentation in detail. Our SkinFormer includes the Kurtosis-guided Statistical Counting Operator (KSCO), the Statistical Texture Fusion Transformer (STFT), and the Statistical Texture Enhance Transformer (STET).

\subsection{Overall Frameworks}
To effectively utilize the statistical texture information of skin lesion images, we propose a statistical texture transformer network (SkinFormer), as shown in Figure~\ref{fig:net}. Our SkinFormer consists of a base network, a Kurtosis-guided Statistical Counting Operator, a Statistical Texture Fusion Transformer, and a Statistical Texture Enhance Transformer. {For the base network, we use U-Net~\cite{Unet}}. For the Kurtosis-guided Statistical Counting Operator, we quantify the input features into multiple levels, and then guide the network to focus on images with large kurtosis values by introducing kurtosis. The statistical texture information of the image is represented by the intensity of each level after quantization.
With the help of Kurtosis-guided Statistical Counting Operators, Statistical Texture Fusion Transformer and Statistical Texture Enhance Transformer are designed.

As shown in Figure~\ref{fig:net}, we employ the deep high-level features of the backbone encoder as the input of the Statistical Texture Fusion Transformer to obtain statistical texture fusion feature $O^{fusion}$ and quantized intensity embedding $S$. $O^{fusion}$ is fed to the decoder, and $S$ is upsampled to twice the original size as the input Q of the local-window self-attention in Statistical Texture Enhance Transformer. The multi-scale features of the backbone encoder are down-sampled to the same scale as the input K, V of the local-window self-attention. {Statistical Texture Enhance Transformer can further enhance texture details and extract texture-related information.}
Finally, the output of the Statistical Texture Enhance Transformer is concatenated with the low-level feature extracted by the backbone decoder, and then the final segmentation prediction map is obtained via a convolutional layer. 

\begin{figure}[htb]
\begin{center}
  \includegraphics[width=0.8\linewidth]{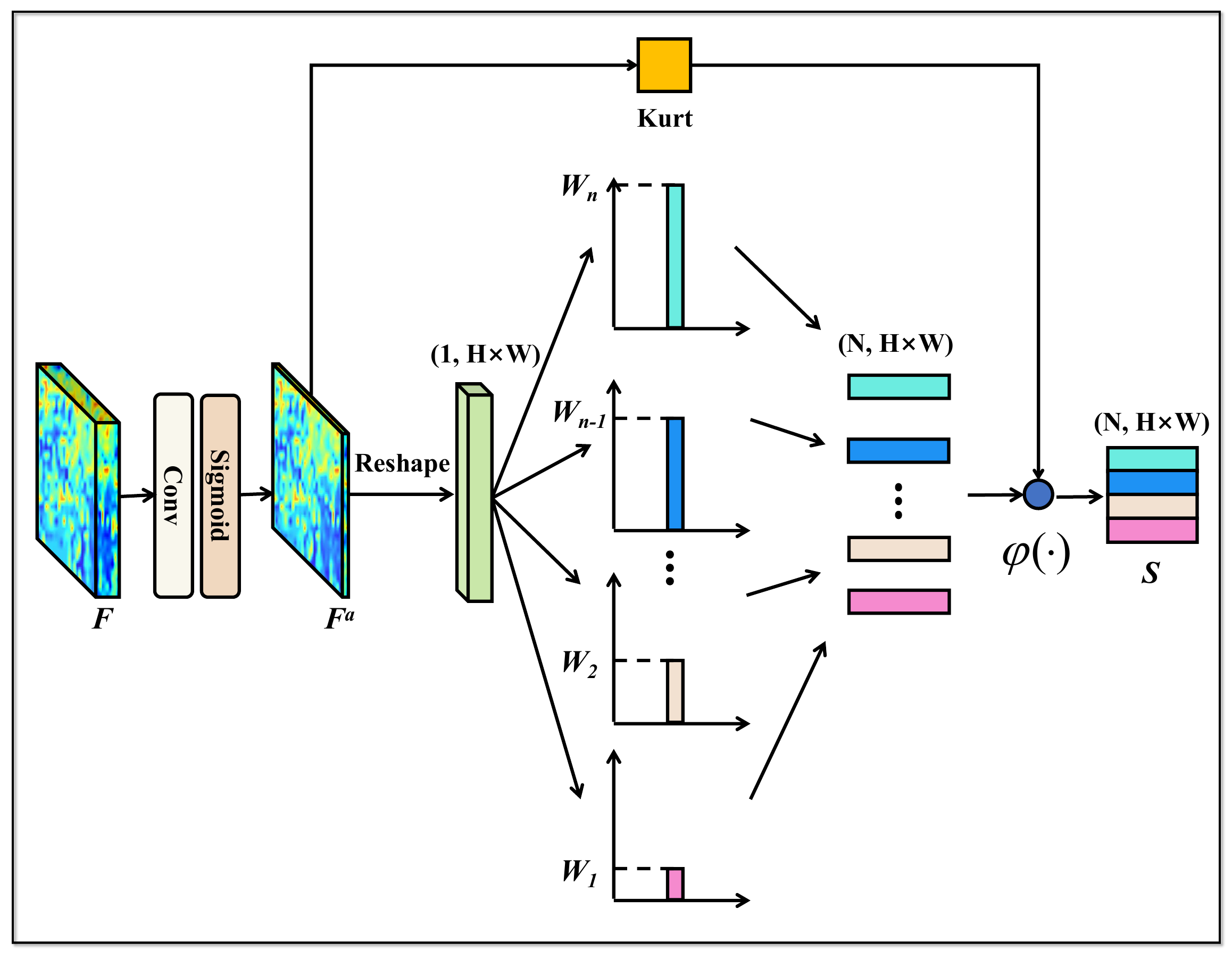}
\end{center}
   \caption{The illustrations of Kurtosis-guided Statistical Counting Operator.
   By adjusting the input feature map $F$'s channel, the feature aggregation map $F^{a}$ is produced. It is then reshaped to have the size of {$\mathbb{R}^{(H\times W)}$}. We quantize the reshaped features into $N$ levels $(W_{1}, W_{2}, ..., W_{N})$. We use the kurtosis of $F^{a}$ as weights and apply $\varphi (\cdot)$ to obtain the quantized intensity embedding $S$.}
\label{fig:ksco}
\end{figure}

\subsection{Kurtosis-guided Statistical Counting Operator}
{Statistical features refer to quantitative descriptors that capture the distribution and variability of pixel intensities in an image. In our paper, these features include metrics such as kurtosis, which measures the "tailedness" of the distribution, providing insights into the texture and structure of the image.}
{The convolution operator is sensitive to local changes in the image and helps to extract local features. However, it cannot efficiently extract statistical textures. Therefore, we propose a Kurtosis-guided Statistical Counting Operator to describe texture representations in a statistical manner.}
{Specifically, the input feature map is denoted as $F \in \mathbb{R}^{C\times H\times W} $}. As shown in Figure~\ref{fig:ksco}, we first get the feature aggregation map $F^{a} \in \mathbb{R}^{1\times H\times W} $ by adjusting the channels via two convolutional layers and sigmoid activation function. Then we reshape the feature aggregation map to {$\mathbb{R}^{(H\times W)}$} size, where the i-th element is denoted as $F^{a}_{i}$. Next, to quantify the statistical texture information, we quantize the reshaped features into $N$ levels $W = [W_{1}, W_{2}, ..., W_{N}]$. The 
n-th level $W_{n}$ is calculated by the following formula:
\begin{align}
W_n = \frac{1}{N} \left [ n\cdot (max(F^{a})-min(F^{a})) + N\cdot min(F^{a}) \right ] 
\end{align}

At the same time, we compute the kurtosis of the above feature aggregation map $F^{a}$. A kurtosis value of 0 indicates that the image completely obeys the normal distribution. We believe that images with a large absolute value of kurtosis have complex contextual information and are worthy of attention. In specific calculations, kurtosis is described as the fourth-order standard moment. {Considering the pixel values of $F^{a}$ as a set of samples $r_{t}, t=1,2,..., (H\times W)$, the kurtosis of $F^{a}$ can be calculated as:}
\begin{align}
K = \frac{1}{T} \sum_{T}^{t=1} (\frac{r_{t} - \bar{r} }{\sigma } )^{4}  
\end{align}

where $\bar{r}$ and $\sigma$ are the mean and standard deviation respectively, which can be expressed as:
\begin{align}
\bar{r}  = \frac{1}{T} \sum_{t=1}^{T} r_{t} ,
\end{align}

\begin{align}
\sigma  = \sqrt{\frac{1}{T-1}\sum_{t=1}^{T}( r_{t} - \bar{r} )^{2} } 
\end{align}

\begin{figure*}[htbp]
\begin{center}  \includegraphics[width=1\linewidth]{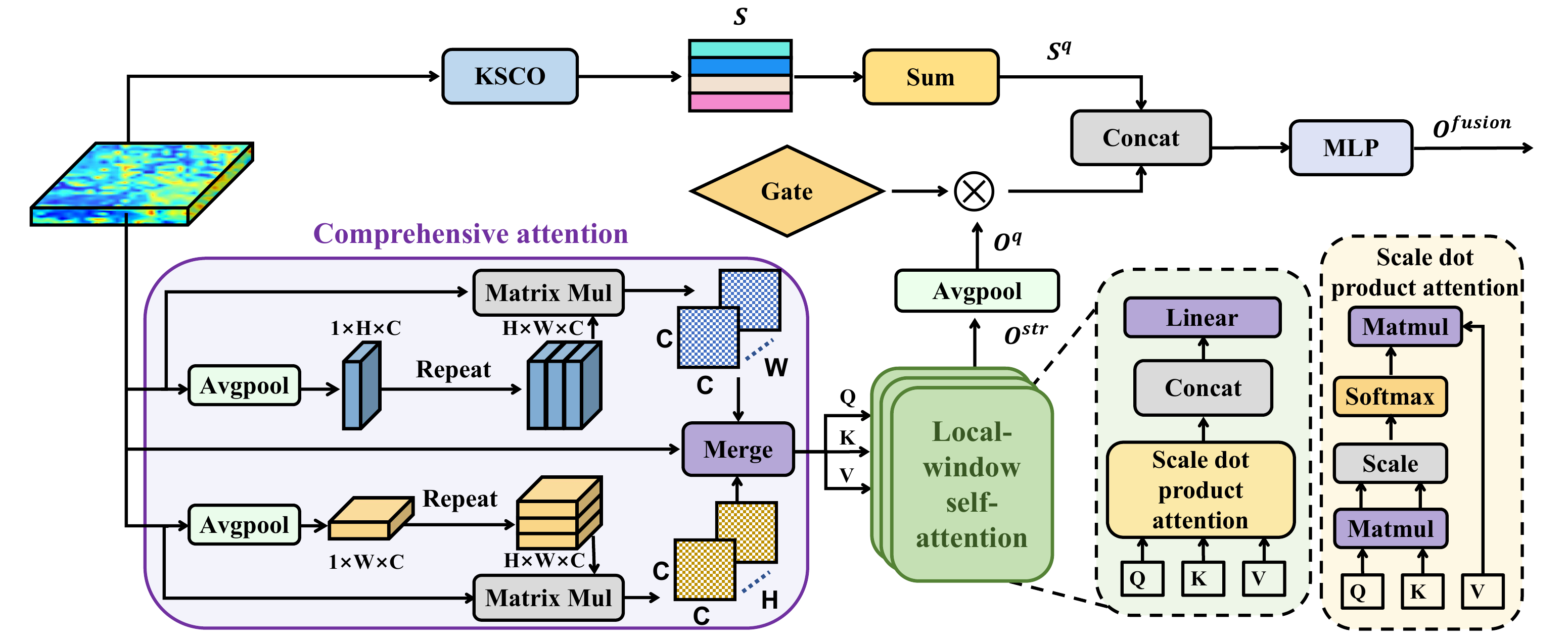}
\end{center}
   \caption{The illustrations of Statistical Texture Fusion Transformer. {The Statistical Texture Fusion Transformer includes Kurtosis-guided Statistical Counting Operator, comprehensive attention, local-window self-attention and gating mechanisms, and is designed to efficiently fuse structural texture information and statistical texture information of images.}}
   
\label{fig:STFT}
\end{figure*}

We use kurtosis as a weight to make the network pay attention to the statistical texture of images with large absolute value of kurtosis. To obtain the quantized intensity of statistical texture information, the quantized intensity embedding {$S\in \mathbb{R}^{N\times (H\times W)} $} is calculated by applying $\varphi (\cdot )$ to the obtained $K, F^{a}_{i},W_{n}$, as shown in Equation~\ref{varphi}.

\begin{align}
S_{simple} = \varphi (K,F^{a}_{i},W_{n}) = \frac{\left | K \right |  }{e^{(1-\left | F^{a}_{i}-W_n \right |) } }
\label{varphi}
\end{align}

In practice, to stabilize the network's training process and expedite operations, our final quantized intensity embedding $S_{i,n}$ is defined as:

\begin{align}
S_{i,n} = 
\left\{\begin{aligned}
&\varphi (K,F^{a}_{i},W_{n}) && ,\left | F^{a}_{i}-W_n \right |< \frac{max(F^{a})-min(F^{a})}{2N} \\
&0 && \text{{,else}}
\end{aligned}\right.
\end{align}

{where $S_{i,n}$ represents the n-th level statistical texture corresponding to the i-th element of $F^{a}$.}
It can be observed that a larger value of $K$ or a smaller value of $1-\left | F^{a}_{i}-W_n \right |$ results in a larger value of $S_{i,n}$. $S_{i,n}$ can reflect the statistical texture quantization level of $F^{a}_{i}$.

\subsection{Statistical Texture Fusion Transformer}
\label{sec:stft}
Statistical texture information and structural texture information are crucial for segmentation tasks. To effectively fuse the structural texture information and statistical texture information of medical images, we design the Statistical Texture Fusion Transformer (STFT). This module incorporates Kurtosis-guided Statistical Counting Operator, local-window self-attention, and gating mechanism to facilitate the performance of segmentation networks.

Specifically, for a given input feature map $F$, we use a Kurtosis-guided Statistical Counting Operator to extract statistical texture information to obtain the corresponding quantized intensity embedding $S$. We further generate the statistical texture quantization aggregation map $S^{q}$ by the following formula:
\begin{equation}
{S^{q} = \sum_{i}^{(H\times W)} S_{i,:} \in \mathbb{R}^{N \times 1}}
\end{equation}
{At the same time, we use comprehensive attention and Local-window self-attention~\cite{yuan2021hrformer} to extract the structural texture information of the feature map $F$,} as shown in Figure~\ref{fig:STFT}.

{
\textbf{Comprehensive attention.}
To improve the representation ability of features, we introduce a comprehensive attention mechanism to simultaneously pay attention to the information in the height and width directions of the features. Unlike previous attention modules~\cite{woo2018cbam} that sequentially apply spatial and channel attention operations to generate attention, comprehensive attention preserves the interaction of information by efficiently computing attention in different directions.

Specifically, we feed the input features $F$ into two parallel branches, each of which contains a global average pooling layer. We extract global features $F^{H} \in \mathbb{R}^{1 \times H \times C}$ and $F^{W} \in \mathbb{R}^{(1 \times W \times C}$ using different pool kernels in width and height directions. Then we repeat $F^{H}$ and $F^{W}$ horizontally and vertically, respectively, and apply matrix multiplication to obtain a horizontal attention map $A^{W}\in \mathbb{R}^{W \times C \times C}$ and a vertical attention map $A^{H}\in \mathbb{R}^{H \times C \times C}$. Finally, in order to pay attention to the information of different dimensions at the same time, $A^{H}$ and $A^{W}$ are multiplied by $F$ respectively and then added and merged into the output.}

\begin{figure*}[htbp]
\begin{center}
  \includegraphics[width=1\linewidth]{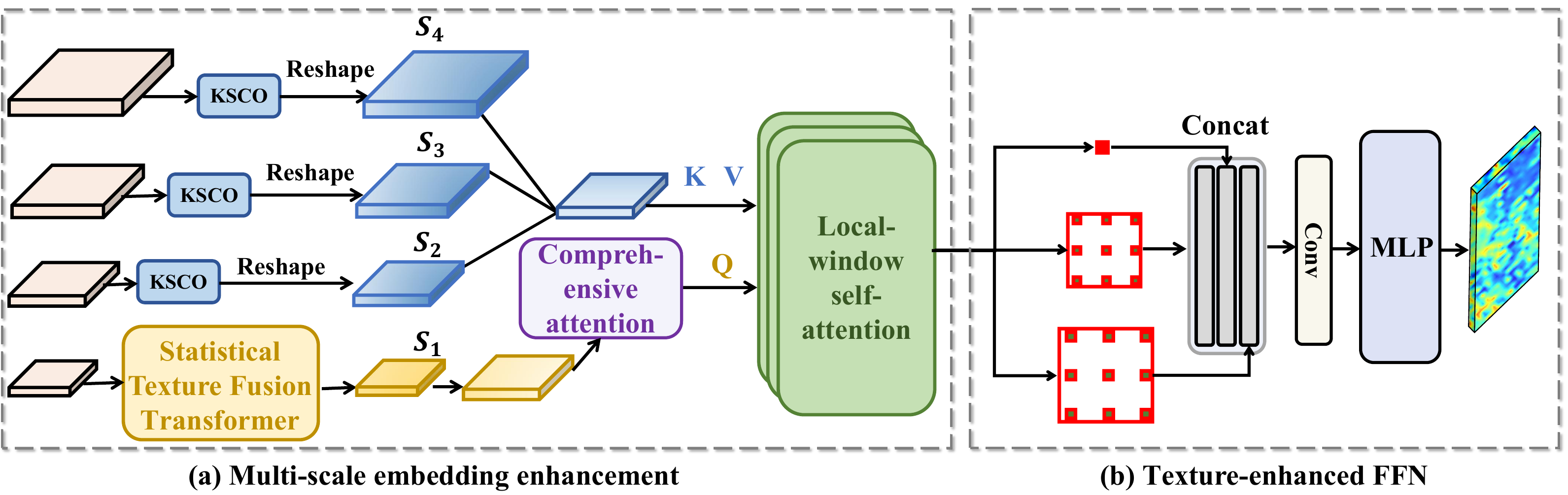}
\end{center}
   \caption{The illustrations of Statistical Texture Enhance Transformer. The Statistical Texture Enhance Transformer is mainly divided into two parts: {(a) Multi-scale embedding enhancement and (b) Texture-enhanced FFN. The former obtains the inputs (Q, K, V) of local window self-attention by enhancing the statistical texture of shallow features and high-level features at different scales, where the statistical texture of high-level features is further enhanced by the comprehensive attention. The latter increases the receptive field by dilated convolution, aiming to enhance the ability of texture representation extraction.}}
  
\label{fig:stet}
\end{figure*}

\textbf{Local-window self-attention.} Local-window self-attention divides the feature map $F \in \mathbb{R}^{HW \times C}$ into a set of non-overlapping small windows of size $K \times K$, and Multi-Head Attention is performed independently within each window. As shown in Figure~\ref{fig:STFT}, Multi-Head Attention consists of linear mapping and scaled Dot-product Attention. For the input $Q, V, K$, they are respectively subjected to $H$ linear transformations to obtain $H$ groups of $Q_{h}, K_{h}, V_{h}, h=1,...,H$. For each set of $Q_{h}, K_{h}, V_{h}$, they are processed by Scaled Dot-product Attention and then connected together. Here $H$ corresponds to the number of heads. For the feature map $F_{p}$ on the p-th window, ($Q_{h}, K_{h}, V_{h}$) correspond to ($F_{p}L_{q}^{h}, F_{p}L_{k}^{h}, F_{p}L_{v}^{h}$). The Multi-Head Attention formula on the p-th window is as follows:

\begin{align}
MHA(F_{p}) = Concat(HA_{1} (F_{p}),...,HA_{H}(F_{p}))\in \mathbb{R}^{K^{2} \times C}
\end{align}

\begin{align}
HA_{h}(F_{p}) = Softmax\left [ \frac{(F_{p}L_{q}^{h})(F_{p}L_{k}^{h})^{T}}{\sqrt{C/H}}) \right ] F_{p}L_{v}^{h} \in \mathbb{R}^{K^{2} \times \frac{C}{H}}
\end{align}

\begin{align}
\hat{F_{p}} = F_{p} + MHA(F_{p})L_{o} \in \mathbb{R}^{K^{2} \times \frac{C}{H}}
\end{align}

Where $L_{q}^{h}\in \mathbb{R}^{\frac{C}{H} \times C}$, $L_{k}^{h}\in \mathbb{R}^{\frac{C}{H} \times C}$, $L_{v}^{h}\in \mathbb{R}^{\frac{C}{H} \times C}$, $L_{o}\in \mathbb{R}^{C \times C}$. We merge all $\hat{F_{p}}$ to compute the structure texture output ${O^{str}}$.

\textbf{Gating mechanism.} To adaptively fuse the structural texture information and statistical texture information, we introduce a gating mechanism. We use a learnable parameter $\alpha$ to represent the degree of fusion of the two texture information. The structure texture output ${O^{str}}$ is further aggregated into a structure texture aggregation map $O^{q}$ through an average pooling. Then $\alpha$ is multiplied by the structure texture aggregation map and concatenated with the statistical texture quantization aggregation map $S^{q}$.
The result obtained is finally fed into MLP to obtain the output $O^{fusion}$. The process can be expressed as:

\begin{align}
O^{q} = avgpool(O^{str})
\end{align}
\begin{align}
O^{fusion} = MLP(Concat(\alpha \cdot O^{q}, S^{q}))
\end{align}


\subsection{Statistical Texture Enhance Transformer}
We further propose a Statistical Texture Enhancement Transformer (STET), which aims to enhance statistical texture-related information on multi-scale feature maps. The shallow feature maps of deep learning networks contain rich detailed texture information. And the fusion of shallow features and high-level features has been verified to be crucial for extracting context and improving accuracy~\cite{Unet}. Therefore, the extraction and integration of statistical texture information in multi-scale features are extremely beneficial to the segmentation performance of the network.

{\textbf{Multi-scale embedding enhancement.}} To enhance the statistical texture-related information of shallow features and high-level features at different scales, the Statistical Texture Enhancement Transformer is designed. As shown in Figure~\ref{fig:stet}, the input of the Statistical Texture Enhancement Transformer is the multi-scale features of the backbone encoder and the quantized intensity embedding $S_{1}$. {We upsample $S$ to twice the original size and then enhance its feature representation by using comprehensive attention, and the result is used as the input Q for the local-window self-attention.} To extract multi-scale statistical texture information, we separately feed the multi-scale features of the backbone encoder to Kurtosis-guided Statistical Counting Operator to extract statistical texture quantized intensity embeddings $S_{2}, S_{3}, S_{4}$. Next, we downsample $S_{2}, S_{3}, S_{4}$ to the same scale as Q as the input K, V of the local-window self-attention.

{\textbf{Texture-enhanced FFN.}} The Statistical Texture Enhance Transformer uses the local-window self-attention introduced in Section~\ref{sec:stft}. Local-window self-attention performs self-attention on windows separately, without cross-window information exchange. {To address this issue and further enhance texture representations, we provide the Texture-enhanced Feed-Forward Network (T-FFN). Specifically, we add three parallel dilated convolutional layers between Local-window self-attention and MLP to enhance the ability of texture representation extraction and information interaction, and their dilation rates are set to 1, 6, and 12, respectively.}

\begin{figure*}[htbp]
\begin{center}
  \includegraphics[width=1\linewidth]{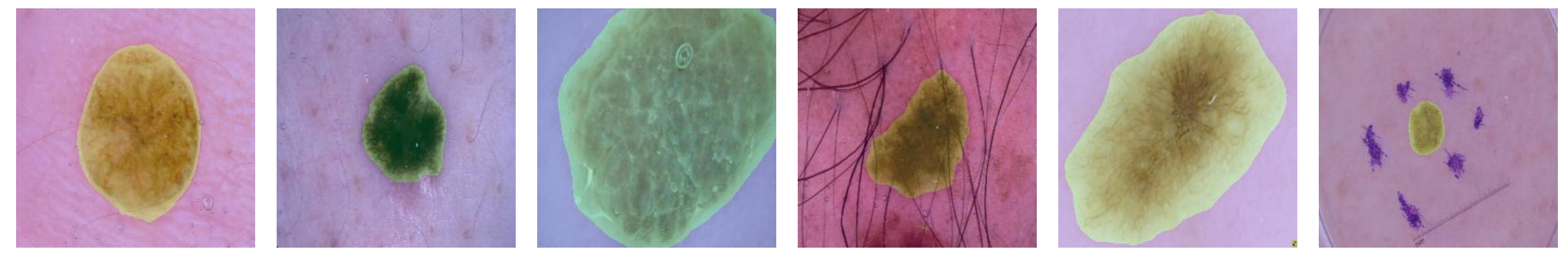}
\end{center}
   \caption{The visualization results of our method on skin lesion images. It can be observed that our SkinFormer can clearly segment the boundaries of the lesions due to learning statistical texture representation via Kurtosis-guided Statistical Counting Operator, Statistical Texture Fusion Transformer, and Statistical Texture Enhance Transformer.}
  
\label{fig:result1}
\end{figure*}

Finally, as shown in Figure~\ref{fig:net}, we concatenate the output of the Statistical Texture Enhance Transformer with the high-level feature map extracted from the decoder, and then feed the concatenated feature map into the convolutional layer to get the final prediction map.
Some visualization results of our SkinFormer on skin lesion images are shown in Figure~\ref{fig:result1}.

\section{Experiments}
\label{exp}
\subsection{Implementation Details and Loss Function} 
Our model is trained with the Adam optimizer. {For training on $ISIC 2018$, SkinFormer took 4 hours using 1 NVIDIA TITAN V. For $ISIC 2018$, we set the batch size to 16 and the iterations to 300 epochs. We set the learning rate to 0.0002 and the weight decay to $10^{- 8}$. We employ a decay strategy to decay the learning rate by 0.5 every 256 epochs. For smaller datasets, we reduce the number of epochs accordingly: 250 epochs for $ISIC 2017$, and 200 epochs for $ISIC 2016$. The decay strategy's starting epoch is also adjusted based on the dataset: epoch 213 for $ISIC 2017$, and epoch 170 for $ISIC 2016$.}

In the experiment, the number of heads $H$ in Local-window self-attention is set to 2, and the window size $K$ in Local-window self-attention is set to 7 for efficiency. {The number of layers $N$ for Kurtosis-guided Statistical Counting Operators in Statistical Texture Fusion Transformer and Statistical Texture Enhance Transformer is set to 256 and 64.} All images are resized to $256\times 256$. We performed the simple data augmentation including vertical flipping and horizontal flipping.

We train the network with the Dice loss~\cite{milletari2016loss} function and test with the best-performing model on the validation set. The Dice loss function can be calculated by the following equation:

\begin{align}
\mathcal L_{\rm dice}=1-\frac{2\sum_{i=1}^{N}(x_{i}{y}_{i})}{\sum_{i=1}^{N}x_{i}^{2}+\sum_{i=1}^{N}{y}_{i}^{2}}.
\end{align}

where $x_{i}$ is the prediction map generated by our method for a given pixel $i$, and ${y}_{i}$ is the corresponding value in the ground truth of the dermoscopy image.


\subsection{Dataset}
We evaluate our SkinFormer on the $ISIC 2016\& PH2$, $ISIC 2017$ and $ISIC 2018$ datasets. {These three datasets are public benchmarks for the skin lesion segmentation.}

{For the $ISIC 2016\& PH2$ dataset, it contains two publicly available skin lesion segmentation datasets: $ISIC 2016$ and $PH2$.
The $ISIC 2016$ dataset contains 900 dermoscopy images, and the $PH2$ dataset includes 200 dermoscopy images. The $ISIC 2016$ can be further subdivided into 727 non-melanoma cases and 173 melanoma cases.
We use the official train-validation split of the $ISIC 2016$ dataset for model learning, and we report testing on 200 samples from the $PH2$.}

For the $ISIC 2017$ dataset, we have the same setting as~\cite{wang2021pr2}, and divide the dataset into a training set containing 2000 dermoscopy images, a validation set containing 150 dermoscopy images, and a testing set containing 600 dermoscopy images set (117 melanoma cases, 393 benign nevi cases, and 90 seborrheic keratosis cases). $ISIC 2017$ is an extension of ISIC 2016, in which dermoscopy images can be further divided into 521 melanoma cases, 386 seborrheic keratosis cases, and 1843 benign nevi cases.

For the $ISIC 2018$ dataset, we adopted the same experimental protocol as in~\cite{gu2020canet}. The $ISIC 2018$ dataset contains 2594 dermoscopy images, which are widely used for skin lesion segmentation, provided by an International Skin Imaging Collaboration challenge. Because the $ISIC 2018$ test set is unannotated, we perform five cross-validations on the $ISIC 2018$ training set for a fair comparison.

\begin{figure*}[thb]
\begin{center}
  \includegraphics[width=0.9\linewidth]{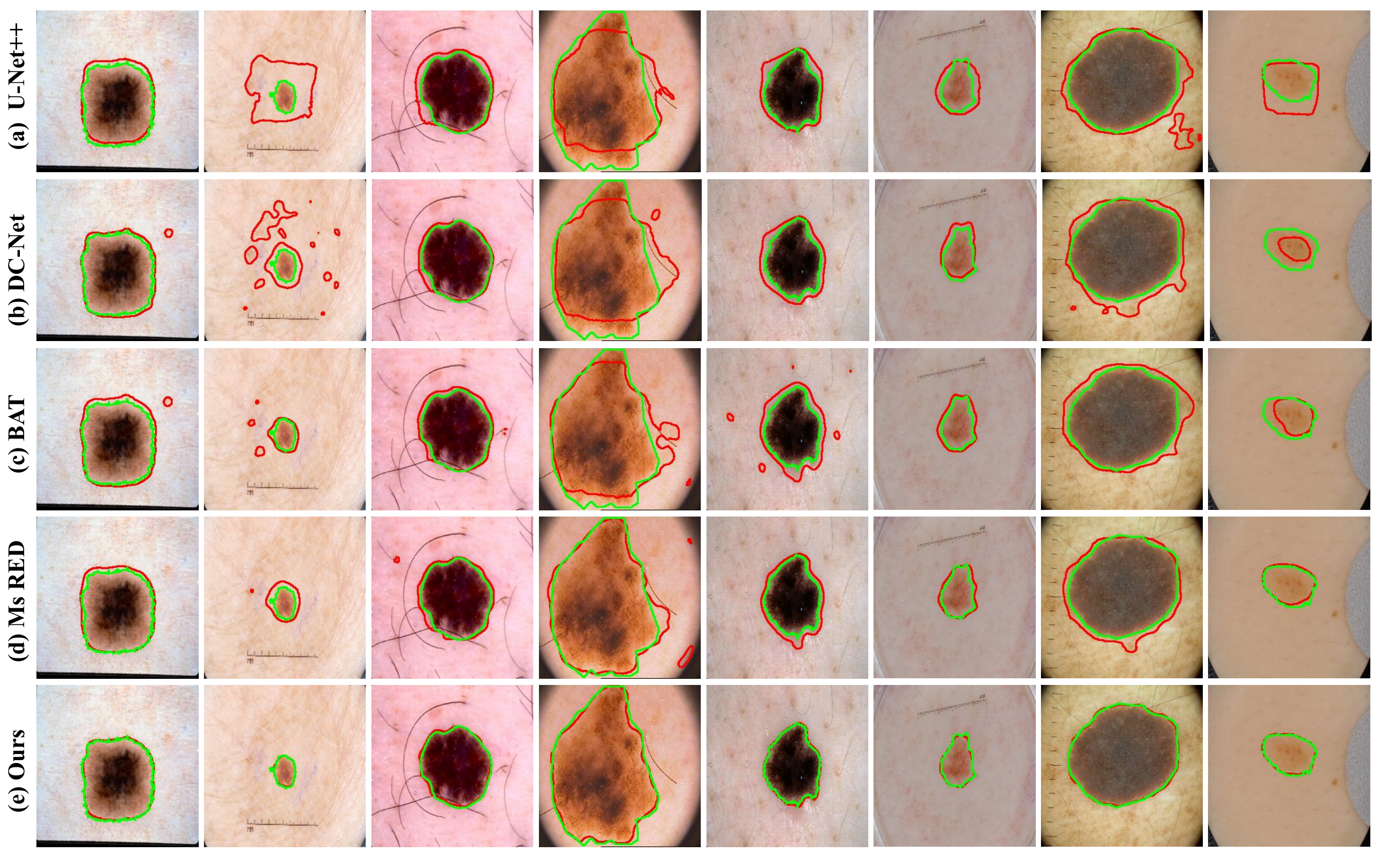}
\end{center}
   \caption{Visual comparison of skin lesion segmentation results produced by our SkinFormer and other methods. The red contours represent the predicted skin lesion segmentation results, and the green contours represent the ground truth. It can be seen that the segmentation results of our SkinFormer have clearer boundaries and details compared with other methods, which are closer to the ground truth.}
\label{fig:result2}
\end{figure*}

\subsection{Evaluation Criteria}
{For the $ISIC 2016\& PH2$ and $ISIC 2018$ datasets, we adopted the Dice coefficient, IoU metric and Hausdorff distance of boundaries (95th percentile; HD95) for lesion segmentation evaluation.}
For the $ISIC 2017$ dataset, we use the same four metrics as~\cite{wang2021pr2}: {Jaccard metric (JA), Dice, segmentation accuracy (AC), and geometric mean (GE)} for the skin lesion segmentation evaluation. GE is the mean of sensitivity and specificity. Dice, mIoU and JA are calculated as:
\begin{align}
Dice = \frac{2TP}{2TP+FP+FN}
\end{align}
\begin{align}
mIoU = \frac{1}{2} (\frac{TP}{TP+FP+FN}+\frac{TN}{TN+FP+FN})
\end{align}
\begin{align}
JA = \frac{TP}{TP+FN+FP}
\end{align}
where $TP$ is the true positives; $TN$ is the number of pixels that correctly segment background pixels, that is, true negatives; $FP$ is false positives; $FN$ is false negatives.

{All experiments are conducted using the test sets provided by each ISIC challenge, adhering to the same setup as competitors and without using external data. We have re-implemented the state-of-the-art methods, including MS RED~\cite{dai2022ms}, and evaluated their performances under identical conditions to ensure a fair comparison.}

\begin{table}[ht]
\caption{{Comparison results on the validation set of $ISIC 2016$ dataset. We report the averaged scores on the $ISIC-2016$. The best results are indicated by the bold value in each column.}}
\label{tab:2016}
\centering
\renewcommand\arraystretch{1} {
\begin{tabular}{lcccc}
\hline
\multicolumn{5}{c}{\textbf{$ISIC 2016-validation$}}  \\ \hline
\multicolumn{1}{c|}{\textbf{Method}} & \textbf{Dice} & \textbf{IoU}& \textbf{HD95}  & \textbf{P-value} (Dice) \\ \hline
\multicolumn{1}{l|}{\textbf{FCN}$_{CVPR’2015}$~\cite{long2015fully}} & 87.2 & 80.3 & 46.4 & $10^{-4}$\\
\multicolumn{1}{l|}{\textbf{U-Net}$_{MICCAI’2015}$~\cite{Unet}} & 87.8 & 80.5 & 45.7 & $10^{-4}$\\
\multicolumn{1}{l|}{\textbf{U-Net++}$_{DLMIA’2018}$~\cite{zhou2018unet++}} & 89.5 & 82.2 & 44.2 & $10^{-4}$\\
\multicolumn{1}{l|}{\textbf{CE-Net}$_{TMI ’2019}$~\cite{CE-Net}} & 90.8 & 84.4 & 31.8& $10^{-3}$\\
\multicolumn{1}{l|}{\textbf{MedT}$_{MICCAI’2021}$~\cite{valanarasu2021medt}} & 88.9 & 81.6 & 31.5& $10^{-3}$\\ 
\multicolumn{1}{l|}{\textbf{DC-Net}$_{MICCAI’2021}$~\cite{xu2021dc}} & 91.0 & 84.6 & 31.1& 0.011\\
\multicolumn{1}{l|}{\textbf{BAT}$_{MICCAI’2021}$~\cite{wang2021bat}} & 91.8 & 85.2 & 30.8 & 0.018\\
\multicolumn{1}{l|}{\textbf{LOBSTER }$_{NN’2022}$~\cite{tartaglione2022loss}} & 90.6 & 83.9 & 31.2 & 0.016\\
\multicolumn{1}{l|}{\textbf{{Ms RED}}$_{MIA’2022}$~\cite{dai2022ms}} & 92.1 & 85.5 & 29.6& 0.026 \\
\multicolumn{1}{l|}{\textbf{Att-SwinU-Net}$_{ISBI’2023}$~\cite{aghdam2023attention}} &  91.3 & 85.1 & 30.6 & 0.028 \\
 \hline
\multicolumn{1}{l|}{\textbf{SkinFormer (Ours)}} & \textbf{93.8} & \textbf{88.0} & \textbf{21.7}& {—}\\ \hline
\end{tabular}}
\end{table}

\subsection{Comparison with Other Methods}

\textbf{Comparisons on the $ISIC 2016\& PH2$ dataset.}
First, we compare our SkinFormer with other SOAT algorithms on the $ISIC 2016\& PH2$ dataset. These methods include {FCN~\cite{long2015fully}, U-Net~\cite{Unet}, U-Net++~\cite{zhou2018unet++}, CE-Net~\cite{CE-Net}, MedT~\cite{valanarasu2021medt}, DC-Net~\cite{xu2021dc}, BAT~\cite{wang2021bat}, LOBSTER~\cite{tartaglione2022loss} and Ms RED~\cite{dai2022ms}.} BAT~\cite{wang2021bat} is a boundary-aware transformer for skin lesion segmentation, achieved the best segmentation performance in recent skin lesion segmentation.
Ms RED~\cite{dai2022ms} is a multi-scale residual encoding and decoding network for skin lesion segmentation.
{We show the comparison results on the validation set of $ISIC 2016$ dataset in Table~\ref{tab:2016},} It is obvious that our SkinFormer achieves the best segmentation performance. Compared with FCN, our SkinFormer improves the Dice metric by 7.6\%, achieving 93.8\%.
Compared with Transformer-based BAT~\cite{wang2021bat}, our SkinFormer improves the IoU metric from {85.5\% to 88.0\%.}
{We also report the results on $PH2$ in Table~\ref{tab:ph2}.} Since images from the $PH2$ dataset are unseen during model training, the satisfactory results demonstrate the excellent generalization ability of our method. Compared with Ms RED~\cite{dai2022ms}, our SkinFormer improves the IoU metric by {2.6\%, achieving 87.4\% on the $PH2$ dataset.}
Furthermore, for the HD95 metric, our model achieves the best performance on both the $ISIC 2016$ validation set and the $PH2$ test set (21.7 and 24.5), which shows that SkinFormer has a promising advantage in handling boundary segmentation.

\begin{table}[tbh]
\caption{{Comparison results on $PH2-test$ dataset for skin lesion segmentation. We report the averaged scores. The best results are indicated by the bold value in each column.}}
\label{tab:ph2}
\centering
        \renewcommand\arraystretch{1.1} {
\begin{tabular}{lcccc}
\hline
\multicolumn{5}{c}{\textbf{$PH2-test$}}  \\ \hline
\multicolumn{1}{c|}{\textbf{Method}} & \textbf{Dice} & \textbf{IoU}& \textbf{HD95}  & \textbf{P-value}(Dice) \\ \hline
\multicolumn{1}{l|}{\textbf{FCN}$_{CVPR’2015}$~\cite{long2015fully}} & 83.3 & 73.8 & 60.7 & $10^{-5}$\\
\multicolumn{1}{l|}{\textbf{U-Net}$_{MICCAI’2015}$~\cite{Unet}} & 83.7 & 74.2 & 59.3 & $10^{-5}$ \\
\multicolumn{1}{l|}{\textbf{U-Net++}$_{DLMIA’2018}$~\cite{zhou2018unet++}} & 89.2 & 82.5 & 45.4& $10^{-4}$\\
\multicolumn{1}{l|}{\textbf{CE-Net}$_{TMI ’2019}$~\cite{CE-Net}} & 90.1 & 83.2 & 34.9& $10^{-3}$\\
\multicolumn{1}{l|}{\textbf{MedT}$_{MICCAI’2021}$~\cite{valanarasu2021medt}} & 88.7 & 80.6 & 36.0& $10^{-4}$\\
\multicolumn{1}{l|}{\textbf{DC-Net}$_{MICCAI’2021}$~\cite{xu2021dc}} & 90.2 & 84.4 & 34.6& $10^{-3}$\\
\multicolumn{1}{l|}{\textbf{BAT}$_{MICCAI’2021}$~\cite{wang2021bat}} & 90.8 & 84.8 & 33.2& 0.011\\
\multicolumn{1}{l|}{\textbf{LOBSTER }$_{NN’2022}$~\cite{tartaglione2022loss}} & 89.5 & 83.6 & 35.1 & 0.010\\
\multicolumn{1}{l|}{\textbf{{Ms RED}}$_{MIA’2022}$~\cite{dai2022ms}} & 91.0 & 85.2 & 32.8& 0.020 \\
\multicolumn{1}{l|}{\textbf{Att-SwinU-Net}$_{ISBI’2023}$~\cite{aghdam2023attention}} &  90.5 & 84.4 & 34.6 & 0.013 \\
 \hline
\multicolumn{1}{l|}{\textbf{SkinFormer (Ours)}} & \textbf{92.7} & \textbf{87.4} & \textbf{24.5}&  {—} \\ \hline
\end{tabular}}
\end{table}

\begin{table}[tbh]
\caption{{Comparison results on $ISIC 2017$ dataset for skin lesion segmentation. We report the averaged scores. The best results are indicated by the bold value in each column.}}
\label{tab:ISIC2017}
\centering
\resizebox{\linewidth}{!}{
\renewcommand\arraystretch{1} {
\begin{tabular}{l|cccccc}
\hline
\multicolumn{1}{c|}{\multirow{2}{*}{\textbf{Method}}} & \multicolumn{2}{c|}{\textbf{Melanoma}} & \multicolumn{2}{c|}{\textbf{Non-Melanoma}} & \multicolumn{2}{c}{\textbf{Overall}} \\ \cline{2-7} 
\multicolumn{1}{c|}{} & JA(\%) & \multicolumn{1}{c|}{AC(\%)} & JA(\%) & \multicolumn{1}{c|}{AC(\%)} & JA(\%) & AC(\%) \\ \hline
\textbf{Team-Yuan}$_{JBHI’2017}$~\cite{yuan2017improving} & 71.2 & 90.0 & 77.8 & 94.2 & 76.5 & 93.4 \\
\textbf{Team-Berseth}$_{ISIC’Challenge}$~\cite{isic2017} & 68.8 & 89.0 & 78.0 & 94.2 & 76.2 & 93.2 \\
\textbf{Team-popleyi}$_{ISIC’Challenge}$~\cite{isic2017} & 69.3 & 89.6 & 77.6 & 94.3 & 76.0 & 93.4 \\
\textbf{Team-Ahn}$_{ISIC’Challenge}$~\cite{isic2017} & 69.1 & 89.6 & 77.5 & 94.3 & 75.8 & 93.4 \\
\textbf{Team-RECOD}$_{ISIC’Challenge}$~\cite{isic2017} & 68.8 & 89.4 & 77.0 & 94.0 & 75.4 & 93.1 \\ 
\textbf{Li et al.}$_{JBHI’2018}$~\cite{li2018dense} & N.A & N.A & N.A & N.A & 76.5 & 93.9 \\
\textbf{Bi et al.}$_{PR’2019}$~\cite{bi2019step} & 72.2 & 90.1 & 79.1 & 95.1 & 77.7 & 94.1 \\
\textbf{Wang et al.}$_{TIP’2019}$~\cite{wang2019bi} & 77.3 & 92.0 & 82.5 & 95.3 & 81.5 & 94.7 \\
\textbf{Xie et al.}$_{TMI’2020}$~\cite{xie2020mutual} & N.A & N.A & N.A & N.A & 80.4 & 94.7 \\
\textbf{Wang et al.}$_{PR’2021}$~\cite{wang2021pr2} & 77.4 & 92.2 & 83.7 & 95.9 & 82.4 & 95.2 \\
\multicolumn{1}{l|}{\textbf{Ms RED}$_{MIA’2022}$~\cite{dai2022ms}} & 77.5 & 92.5 & 83.9 & 96.0 & 82.5 & 95.3  \\ \hline
\textbf{SkinFormer (Ours)} & \textbf{78.0} & \textbf{93.1} & \textbf{85.7} & \textbf{96.8} & \textbf{84.2} & \textbf{96.1} \\ \hline
\end{tabular}}}
\end{table}

\textbf{Comparisons on the $ISIC 2017$ dataset.}
To fully confirm that our SkinFormer is effective, we also compare our SkinFormer with other SOTA methods on the $ISIC 2017$ dataset. We follow the comparison method of Wang et al.~\cite{wang2021pr2} without introducing any other external training data. {Table~\ref{tab:ISIC2017} reports the performance of our method on the Jaccard metric (JA)} and segmentation accuracy (AC), respectively.
It can be observed that our SkinFormer achieves the best JA metric of {84.2\%, which is 2.1\%} higher than the best reported result of method~\cite{dai2022ms}. And our method increases the AC metric of \cite{dai2022ms} from {95.3\% to 96.1\%}.
This benefits from the efficient use of statistical texture information by our SkinFormer. 
Some visual examples comparing with different methods are shown in Figure~\ref{fig:result2}.
It can be observed that the segmentation results of our SkinFormer have clearer boundaries and details compared to other SOAT methods, proving the effectiveness of our SkinFormer.
{As shown in Figure~\ref{fig:result2}, although our method achieves excellent segmentation results, the part with blurred boundaries needs to be further improved. In extreme cases where the boundaries of skin lesions are highly similar to their surroundings, our method may encounter difficulties in accurately segmenting the lesion.}

\begin{table}[tbh]
\caption{{Comparison results on $ISIC 2018$ dataset for skin lesion segmentation. We report the averaged scores in the table. The best results are indicated by the bold value in each column.}}
\label{tab:2018}
\centering
\resizebox{\linewidth}{!}{
\renewcommand\arraystretch{1.1} {
\begin{tabular}{lcccc}
\hline
\multicolumn{5}{c}{\textbf{$ISIC 2018$}} \\ \hline
\multicolumn{1}{c|}{\textbf{Method}} & \textbf{JA(\%)} &\textbf{Dice(\%)} & \textbf{IoU(\%)} & \textbf{HD95} \\ \hline
\multicolumn{1}{l|}{\textbf{DeepLabv3}$_{ECCV’2018}$~\cite{chen2018DeepLabV3}} &81.2±0.4 & 88.5±0.7 & 80.7±0.5 & 36.9±0.6  \\
\multicolumn{1}{l|}{\textbf{U-Net++}$_{DLMIA’2018}$~\cite{zhou2018unet++}} & 81.0±0.5& 87.8±0.6 & 80.6±0.4 & 41.2±0.3  \\
\multicolumn{1}{l|}{\textbf{CE-Net}$_{TMI’2019}$~\cite{CE-Net}} & 82.1±0.4 & 89.0±0.3 & 81.4±0.3 & 35.1±0.4 \\
\multicolumn{1}{l|}{\textbf{MedT}$_{MICCAI’2021}$~\cite{valanarasu2021medt}} & 79.7±0.7 & 86.1±0.6 & 78.2±0.8 & 50.3±0.6 \\
\multicolumn{1}{l|}{\textbf{MCTrans}$_{MICCAI’2021}$~\cite{ji2021multi}} &80.3±0.5 & 86.8±0.6 & 79.6±0.6 & 45.7±0.5 \\
\multicolumn{1}{l|}{\textbf{DC-Net}$_{MICCAI’2021}$~\cite{xu2021dc}} & 82.3±0.6 & 90.7±0.5 & 83.2±0.4  & 33.8±0.4  \\
\multicolumn{1}{l|}{\textbf{BAT}$_{MICCAI’2021}$~\cite{wang2021bat}} &82.7±0.6 & 91.2±0.6 & 84.3±0.7 & 32.4±0.5 \\
\multicolumn{1}{l|}{\textbf{LOBSTER }$_{NN’2022}$~\cite{tartaglione2022loss}} & 81.8±0.4 & 90.0±0.3 & 82.9±0.2 & 34.2±0.4 \\
\multicolumn{1}{l|}{\textbf{Ms RED}$_{MIA’2022}$~\cite{dai2022ms}} & 84.0±0.4& 91.4±0.5  & 84.5±0.5  & 31.5±0.3  \\
\multicolumn{1}{l|}{\textbf{Att-SwinU-Net}$_{ISBI’2023}$~\cite{aghdam2023attention}} & 83.5±0.6&  91.3±0.5 & 85.1±0.7 & 30.6±0.5  \\
 \hline
\multicolumn{1}{l|}{\textbf{SkinFormer (Ours)}} & \textbf{87.9±0.2} &\textbf{93.2±0.3} & \textbf{87.6±0.3} & \textbf{21.9±0.3} \\ \hline
\end{tabular}}}
\end{table}

\textbf{Comparisons on the $ISIC 2018$ dataset.}
Further, we apply our method to the $ISIC 2018$ dataset. We compare our SkinFormer with other recent methods on this dataset. These methods include classic DeepLabv3~\cite{chen2018DeepLabV3}, attention-based U-Net++~\cite{zhou2018unet++}, CE-Net~\cite{CE-Net}, and state-of-the-art methods (MedT~\cite{valanarasu2021medt}, MCTrans~\cite{ji2021multi}, DC-Net~\cite{xu2021dc}, Ms RED~\cite{dai2022ms} and BAT~\cite{wang2021bat}) based on hybrid architecture of CNN and transformer. Table~\ref{tab:2018} shows the performance comparison of our SkinFormer and other advanced methods on $ISIC 2018$ dataset. {It can be observed from the table that our method achieves the best performance on the Jaccard score, which is attributed to the effective extraction and fusion of statistical texture representation. Moreover, it can be observed from the table that the overall reliability of our method is relatively high.} Compared to other segmentation methods, our SkinFormer achieves the highest skin lesion segmentation accuracy with a Dice metric of {93.2\%.}
Compared with Ms RED~\cite{dai2022ms}, our method improves the IoU metric by {3.7\%, achieving 87.6\%.}
Compared with transformer-based BAT~\cite{wang2021bat}, our method also achieves improvements by {2.2\% and 3.9\%} on Dice metric and IoU metric, respectively, verifying the effectiveness of our method. 
{At the same time, our method also achieves the lowest HD95 value, indicating the superiority of our method in boundary segmentation.}
Our SkinFormer can achieve excellent performance due to learning statistical texture representation via Kurtosis-guided Statistical Counting Operator, Statistical Texture Fusion Transformer, and Statistical Texture Enhance Transformer.

Furthermore, we compared our SkinFormer with other methods by performing the Wilcoxon rank-sum test for statistical testing. {The P-values in Table~\ref{tab:2016} and Table~\ref{tab:ph2} show that our SkinFormer achieves a significant improvement on the Dice metric at the 5\% level ($p < 0.05$ for all). We also conduct additional experiments using the HAM10000 dataset~\cite{tschandl2018ham10000}. Our approach demonstrated superior performance compared to the methods discussed in the paper. Specifically, our method achieved a Dice score of 91.6, which is higher than the 88.4 achieved by the Ms RED~\cite{dai2022ms} method.}

\subsection{Ablation Study}
\subsubsection{Ablation Study of Proposed Components}
To evaluate the effectiveness of our SkinFormer and each component in our framework, we first conduct ablation experiments on {$ISIC 2016$} dataset and $ISIC 2018$ dataset. The baseline adopts our implemented U-Net.
As shown in Table~\ref{tab:abl1}, the results show that both the designed Statistical Texture Fusion Transformer (STFT) and Statistical Texture Enhance Transformer (STET) with the help of the Kurtosis-guided Statistical Counting Operator (KSCO) are crucial for improving the skin lesion segmentation accuracy.
Compared to the baseline, after applying the {STFT}, the Dice metric increases from {87.8\% to 93.8\% on $ISIC 2016$ validation set.}
On the basis of applying STFT, after we further applied {STET}, the Dice metric increases from {91.2\% to 93.2\%} on $ISIC 2018$.

\begin{figure}[th]
\begin{center}
  \includegraphics[width=1\linewidth]{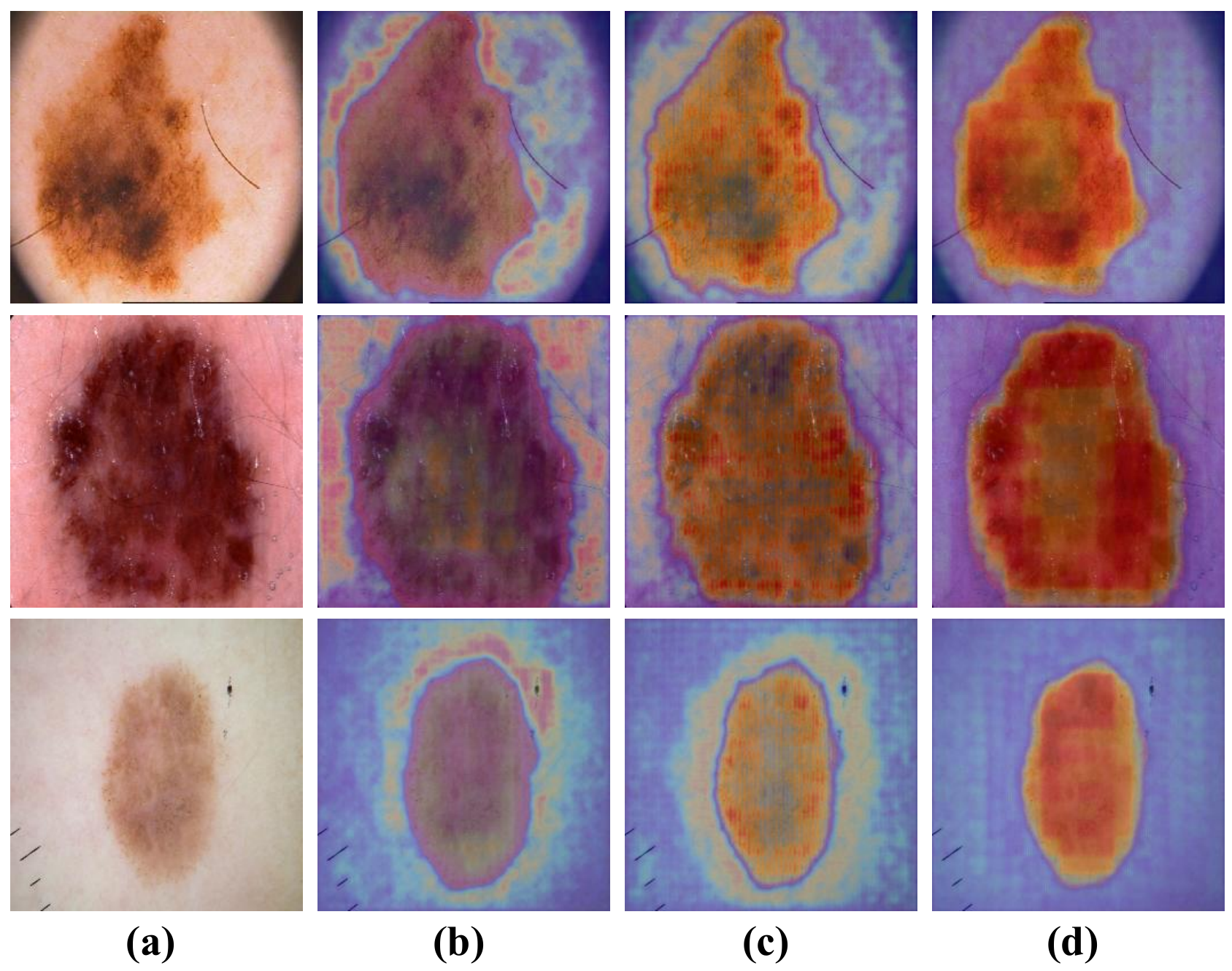}
\end{center}
   \caption{{Comparison of visual heat maps for the last layer of the decoder. (a) Original images, (b) heat maps of baseline, (c) heat maps after applying Statistical Texture Fusion Transformer, (d) heat maps after applying Statistical Texture Enhance Transformer (i.e. our SkinFormer).}}
\label{fig:end}
\end{figure}

\begin{table}[tbh]
\caption{{Ablation experiments of different proposed components on $ISIC 2016$ validation set and $ISIC 2018$ dataset. Here \CheckmarkBold indicates that this component is applied. The best results are indicated by the bold value in each column.}}
\label{tab:abl1}
\centering
\renewcommand\arraystretch{1.1} {
\begin{tabular}{cl|cc|cc}
\hline
\multicolumn{2}{c|}{\textbf{Variants}} & \multicolumn{2}{c|}{\textbf{$ISIC 2016-validation$}} & \multicolumn{2}{c}{\textbf{$ISIC 2018$}} \\ \hline
\textbf{STFT} & \multicolumn{1}{c|}{\textbf{STET}} & \multicolumn{1}{l}{\textbf{Dice(\%)}} & \multicolumn{1}{l|}{\textbf{mIoU(\%)}} & \multicolumn{1}{l}{\textbf{Dice(\%)}} & \multicolumn{1}{l}{\textbf{mIoU(\%)}} \\ \hline
\textbf{} & \textbf{} & 87.8 & 80.5 & 87.2 & 80.1 \\
\CheckmarkBold & \textbf{} & 90.7 & {84.2} & 91.2 & 86.4 \\
\CheckmarkBold & \CheckmarkBold & \textbf{93.8} & \textbf{88.0} & \textbf{93.2} & \textbf{87.6} \\ \hline
\end{tabular}}
\end{table}

\begin{table}[tbh]
\caption{{Ablation experiments of different proposed components on $ISIC 2017$ dataset. Here \CheckmarkBold indicates that this component is applied. The best results are indicated by the bold value in each column.}}
\label{tab:abl2}
\centering
\resizebox{\linewidth}{!}{
\begin{tabular}{cl|cccc|cc}
\hline
\multicolumn{2}{c|}{\textbf{Variants}} & \multicolumn{5}{c}{\textbf{$ISIC 2017$}} \\ \hline
\textbf{STFT} & \multicolumn{1}{c|}{\textbf{STET}} & \textbf{JA(\%)} & \textbf{AC(\%)} & \textbf{Dice(\%)} & \textbf{GM(\%)} & \textbf{Para(M)}& \textbf{FLOPs}\\ \hline
 &  & 75.3 & 92.9 & 84.0 & 87.1 & 11.5 &3.5G\\
\CheckmarkBold & \textbf{} & 82.4 & 93.8 & 90.3 & 91.0 & 
 11.9 &3.8G\\
\CheckmarkBold & \CheckmarkBold &\textbf{84.2} & \textbf{96.1} & \textbf{92.0} & \textbf{92.9} & 12.4 &4.1G \\ \hline
\end{tabular}}
\end{table}

\begin{table}[tbh]
\caption{{Ablation experiments of Statistical Texture Fusion Transformer on $ISIC 2016$ dataset and $ISIC 2018$ dataset. Here \CheckmarkBold indicates that this component is applied. The best results are indicated by the bold value in each column.}}
\label{tab:abl3}
\centering
\resizebox{\linewidth}{!}{
\renewcommand\arraystretch{1} {
\begin{tabular}{cll|cc|cc} 
\hline
\multicolumn{3}{c|}{\textbf{Variants}} & \multicolumn{2}{c|}{\textbf{$ISIC 2016-validation$}} & \multicolumn{2}{c}{\textbf{$ISIC 2018$}} \\ 
\hline
\textbf{KSCO} & \multicolumn{1}{c}{\textbf{Gating}} & \textbf{CA} & \multicolumn{1}{l}{\textbf{Dice(\%)}} & \multicolumn{1}{l|}{\textbf{mIoU(\%)}} & \multicolumn{1}{l}{\textbf{Dice(\%)}} & \multicolumn{1}{l}{\textbf{mIoU(\%)}} \\ 
\hline
  &  &  & 87.8 & 80.5 & 87.2 & 80.1 \\
 \CheckmarkBold &  &  & 88.7 & 81.9 & 88.9 & 82.3 \\
 \CheckmarkBold & \CheckmarkBold &  & 89.5 & 83.1 & 90.4 & 85.6 \\
\CheckmarkBold & \CheckmarkBold & \CheckmarkBold & \textbf{90.7} & \textbf{84.2} & \textbf{91.2} & \textbf{86.4} \\
\hline
\end{tabular}}}
\end{table}

\begin{table}[t]
\caption{{Ablation experiments of Statistical Texture Enhance Transformer on $ISIC 2016$ and $ISIC 2018$ dataset. Here \CheckmarkBold indicates that this component is applied. The best results are indicated by the bold value in each column.}}
\centering
\renewcommand\arraystretch{1} {
\begin{tabular}{cl|cc|cc}
\hline
\multicolumn{2}{c|}{\textbf{Variants}} & \multicolumn{2}{c|}{\textbf{$ISIC 2016-validation$}} & \multicolumn{2}{c}{\textbf{$ISIC 2018$}} \\ \hline
\textbf{MSEE} & \multicolumn{1}{c|}{\textbf{T-FFN}} & \multicolumn{1}{l}{\textbf{Dice(\%)}} & \multicolumn{1}{l|}{\textbf{mIoU(\%)}} & \multicolumn{1}{l}{\textbf{Dice(\%)}} & \multicolumn{1}{l}{\textbf{mIoU(\%)}} \\ \hline
\textbf{} & \textbf{} & 87.8 & 80.5 & 87.2 & 80.1 \\
\CheckmarkBold & \textbf{} & 89.4 & 82.9 & 89.7 & 84.5 \\
\CheckmarkBold & \CheckmarkBold & \textbf{91.0} & \textbf{85.2} & \textbf{91.4} & \textbf{86.5} \\ \hline
\end{tabular}}
\label{tab:abl4}
\end{table}

We also conduct ablation experiments with the same settings on the $ISIC 2017$ dataset to demonstrate the effectiveness of our SkinFormer. The results in Table~\ref{tab:abl2} show that both Statistical Texture Fusion Transformer and Statistical Texture Enhance Transformer help improve the network's performance for skin lesion segmentation. Compared with the baseline, our method improves the Dice metric by {9.5\%.}
This verifies that our SkinFormer can more accurately segment the skin lesions with different shapes, colors and textures.
{In Table~\ref{tab:abl2} we report the number of parameters and FLOPs after applying our components when the input size is $256\times 256$. KSCO only contains one convolutional layer, which does not bring the burden of computation. Thanks to our lightweight attention module design, the increase in the number of parameters is minimal.} Therefore, the FLOPs after applying Statistical Texture Fusion Transformer only increase by 8.6\%, and the increase in FLOPs after applying Statistical Texture Enhance Transformer mainly comes from the processing of high-resolution features. The results show a substantial improvement in segmentation performance with an acceptable increase in computational overhead.
{Furthermore, we visualize the comparison of the heat maps in Figure~\ref{fig:end}, demonstrating the effectiveness of the proposed Statistical Texture Fusion Transformer and Statistical Texture Enhance Transformer.}

\subsubsection{Ablation Study of Statistical Texture Fusion Transformer}

In this section, to explore the impact of {KSCO (Kurtosis-guided Statistical Counting Operator), CA (comprehensive attention) and gating mechanism (Gating)} in our proposed Statistical Texture Fusion Transformer, we design the ablation experiments shown in Table~\ref{tab:abl3} on $ISIC 2016$ dataset and $ISIC 2018$ dataset. The baseline here is the U-Net we implemented. Experiments show that our KSCO can effectively extract statistical texture information, which has a huge impact on improving segmentation performance. The gating mechanism adaptively adjusts the fusion degree of structural texture information and statistical texture information, which is crucial to the segmentation results. {In addition, comprehensive attention effectively enhances the feature representation ability and improves the segmentation performance.}
In summary, benefiting from the effective fusion of structural texture representation and statistical texture representation by the Statistical Texture Fusion Transformer, the segmentation performance of skin lesion images is improved.

\subsubsection{Ablation Study of Statistical Texture Enhance Transformer}

In this section, we further explore the role of different components in the Statistical Texture Enhance Transformer. The results of ablation experiments on $ISIC 2016$ dataset and $ISIC 2018$ dataset are shown in Table~\ref{tab:abl4}. The baseline here is our implemented U-Net. Experimental results show that {Multi-scale embedding enhancement (MSEE) and Texture-enhanced FFN (T-FFN)} are crucial for enhancing multi-scale statistical texture extraction and boosting the segmentation accuracy.

\section{Conclusions}
In this paper, we provide a transformer network (SkinFormer) that extracts and fuses statistical texture representations for accurate segmentation of skin lesion images. First, we present a Kurtosis-guided Statistical Counting Operator to quantify input features and use kurtosis to guide the network to focus on the images with large kurtosis values. Then, we design the Statistical Texture Fusion Transformer to effectively fuse structural texture information and statistical texture information. To further enhance the segmentation performance, we also design the Statistical Texture Enhance Transformer to enhance the statistical texture details of multi-scale features. {Our SkinFormer achieves a Dice score of 93.2\% on the $ISIC 2018$ dataset.} Comparisons with existing advanced methods validate that our SkinFormer achieves SOTA performance on the $ISIC 2016\& PH2$ datasets, $ISIC 2017$ datasets and $ISIC 2018$ datasets. Extensive ablation experiments on skin lesion datasets illustrate the effectiveness of our proposed components.
In the future, {it would be interesting to apply our SkinFormer to other image modalities such as 3D images.}

\section{Acknowledgements}
This work was supported by the National Natural Science Foundation of China (Nos. 62171321,  62071157, 62162044, 62102414, and 62365014).

\bibliographystyle{ieeetran}
\bibliography{main}

\begin{thebibliography}{10}
\providecommand{\url}[1]{#1}
\csname url@samestyle\endcsname
\providecommand{\newblock}{\relax}
\providecommand{\bibinfo}[2]{#2}
\providecommand{\BIBentrySTDinterwordspacing}{\spaceskip=0pt\relax}
\providecommand{\BIBentryALTinterwordstretchfactor}{4}
\providecommand{\BIBentryALTinterwordspacing}{\spaceskip=\fontdimen2\font plus
\BIBentryALTinterwordstretchfactor\fontdimen3\font minus
  \fontdimen4\font\relax}
\providecommand{\BIBforeignlanguage}[2]{{%
\expandafter\ifx\csname l@#1\endcsname\relax
\typeout{** WARNING: IEEEtran.bst: No hyphenation pattern has been}%
\typeout{** loaded for the language `#1'. Using the pattern for}%
\typeout{** the default language instead.}%
\else
\language=\csname l@#1\endcsname
\fi
#2}}
\providecommand{\BIBdecl}{\relax}
\BIBdecl

\bibitem{liu2021fcp}
Y.~Liu, J.~Zhou, L.~Liu, Z.~Zhan, Y.~Hu, Y.~Q. Fu, and H.~Duan, ``Fcp-net: A
  feature-compression-pyramid network guided by game-theoretic interactions for
  skin lesion segmentation,'' \emph{IEEE Transactions on Medical Imaging},
  2021.

\bibitem{xie2016melanoma}
F.~Xie, H.~Fan, Y.~Li, Z.~Jiang, R.~Meng, and A.~Bovik, ``Melanoma
  classification on dermoscopy images using a neural network ensemble model,''
  \emph{IEEE transactions on medical imaging}, vol.~36, no.~3, pp. 849--858,
  2016.

\bibitem{gessert2019skin_tbme}
N.~Gessert, T.~Sentker, F.~Madesta, R.~Schmitz, H.~Kniep, I.~Baltruschat,
  R.~Werner, and A.~Schlaefer, ``Skin lesion classification using cnns with
  patch-based attention and diagnosis-guided loss weighting,'' \emph{IEEE
  Transactions on Biomedical Engineering}, vol.~67, no.~2, pp. 495--503, 2019.

\bibitem{long2015fully}
J.~Long, E.~Shelhamer, and T.~Darrell, ``Fully convolutional networks for
  semantic segmentation,'' in \emph{Proceedings of the IEEE conference on
  computer vision and pattern recognition}, 2015, pp. 3431--3440.

\bibitem{bi2019step}
L.~Bi, J.~Kim, E.~Ahn, A.~Kumar, D.~Feng, and M.~Fulham, ``Step-wise
  integration of deep class-specific learning for dermoscopic image
  segmentation,'' \emph{Pattern recognition}, vol.~85, pp. 78--89, 2019.

\bibitem{Unet}
O.~Ronneberger, P.~Fischer, and T.~Brox, ``U-net: Convolutional networks for
  biomedical image segmentation,'' in \emph{International Conference on Medical
  image computing and computer-assisted intervention}.\hskip 1em plus 0.5em
  minus 0.4em\relax Springer, 2015, pp. 234--241.

\bibitem{chen2018DeepLabV3}
L.-C. Chen, Y.~Zhu, G.~Papandreou, F.~Schroff, and H.~Adam, ``Encoder-decoder
  with atrous separable convolution for semantic image segmentation,'' in
  \emph{Proceedings of the European conference on computer vision (ECCV)},
  2018, pp. 801--818.

\bibitem{haralick1973textural}
R.~M. Haralick, K.~Shanmugam, and I.~H. Dinstein, ``Textural features for image
  classification,'' \emph{IEEE Transactions on systems, man, and cybernetics},
  no.~6, pp. 610--621, 1973.

\bibitem{castleman1996digital}
K.~R. Castleman, \emph{Digital image processing}.\hskip 1em plus 0.5em minus
  0.4em\relax Prentice Hall Press, 1996.

\bibitem{yuan2021hrformer}
Y.~Yuan, R.~Fu, L.~Huang, W.~Lin, C.~Zhang, X.~Chen, and J.~Wang, ``Hrformer:
  High-resolution transformer for dense prediction,'' \emph{arXiv preprint
  arXiv:2110.09408}, 2021.

\bibitem{bi2017dermoscopic_tbme}
L.~Bi, J.~Kim, E.~Ahn, A.~Kumar, M.~Fulham, and D.~Feng, ``Dermoscopic image
  segmentation via multistage fully convolutional networks,'' \emph{IEEE
  Transactions on Biomedical Engineering}, vol.~64, no.~9, pp. 2065--2074,
  2017.

\bibitem{glaister2014segmentation_tbme}
J.~Glaister, A.~Wong, and D.~A. Clausi, ``Segmentation of skin lesions from
  digital images using joint statistical texture distinctiveness,'' \emph{IEEE
  transactions on biomedical engineering}, vol.~61, no.~4, pp. 1220--1230,
  2014.

\bibitem{sadri2012segmentation_tbme}
A.~R. Sadri, M.~Zekri, S.~Sadri, N.~Gheissari, M.~Mokhtari, and F.~Kolahdouzan,
  ``Segmentation of dermoscopy images using wavelet networks,'' \emph{IEEE
  Transactions on Biomedical Engineering}, vol.~60, no.~4, pp. 1134--1141,
  2012.

\bibitem{tartaglione2022loss}
E.~Tartaglione, A.~Bragagnolo, A.~Fiandrotti, and M.~Grangetto, ``Loss-based
  sensitivity regularization: towards deep sparse neural networks,''
  \emph{Neural Networks}, vol. 146, pp. 230--237, 2022.

\bibitem{zhou2022h}
X.~Zhou, X.~Nie, Z.~Li, X.~Lin, E.~Xue, L.~Wang, J.~Lan, G.~Chen, M.~Du, and
  T.~Tong, ``H-net: A dual-decoder enhanced fcnn for automated biomedical image
  diagnosis,'' \emph{Information Sciences}, vol. 613, pp. 575--590, 2022.

\bibitem{yuan2017automatic}
Y.~Yuan, M.~Chao, and Y.-C. Lo, ``Automatic skin lesion segmentation using deep
  fully convolutional networks with jaccard distance,'' \emph{IEEE transactions
  on medical imaging}, vol.~36, no.~9, pp. 1876--1886, 2017.

\bibitem{mirikharaji2018star}
Z.~Mirikharaji and G.~Hamarneh, ``Star shape prior in fully convolutional
  networks for skin lesion segmentation,'' in \emph{International Conference on
  Medical Image Computing and Computer-Assisted Intervention}.\hskip 1em plus
  0.5em minus 0.4em\relax Springer, 2018, pp. 737--745.

\bibitem{zhou2018unet++}
Z.~Zhou, M.~M. Rahman~Siddiquee, N.~Tajbakhsh, and J.~Liang, ``Unet++: A nested
  u-net architecture for medical image segmentation,'' in \emph{Deep learning
  in medical image analysis and multimodal learning for clinical decision
  support}.\hskip 1em plus 0.5em minus 0.4em\relax Springer, 2018, pp. 3--11.

\bibitem{tang2019efficient}
P.~Tang, Q.~Liang, X.~Yan, S.~Xiang, W.~Sun, D.~Zhang, and G.~Coppola,
  ``Efficient skin lesion segmentation using separable-unet with stochastic
  weight averaging,'' \emph{Computer methods and programs in biomedicine}, vol.
  178, pp. 289--301, 2019.

\bibitem{CE-Net}
Z.~{Gu}, J.~{Cheng}, H.~{Fu}, K.~{Zhou}, H.~{Hao}, Y.~{Zhao}, T.~{Zhang},
  S.~{Gao}, and J.~{Liu}, ``Ce-net: Context encoder network for 2d medical
  image segmentation,'' \emph{IEEE Transactions on Medical Imaging}, vol.~38,
  no.~10, pp. 2281--2292, 2019.

\bibitem{qamar2021dense}
S.~Qamar, P.~Ahmad, and L.~Shen, ``Dense encoder-decoder--based architecture
  for skin lesion segmentation,'' \emph{Cognitive Computation}, vol.~13, no.~2,
  pp. 583--594, 2021.

\bibitem{wu2022fat}
H.~Wu, S.~Chen, G.~Chen, W.~Wang, B.~Lei, and Z.~Wen, ``Fat-net: Feature
  adaptive transformers for automated skin lesion segmentation,'' \emph{Medical
  image analysis}, vol.~76, p. 102327, 2022.

\bibitem{ruan2022mew}
J.~Ruan, M.~Xie, S.~Xiang, T.~Liu, and Y.~Fu, ``Mew-unet: Multi-axis
  representation learning in frequency domain for medical image segmentation,''
  \emph{arXiv preprint arXiv:2210.14007}, 2022.

\bibitem{dosovitskiy2020image}
A.~Dosovitskiy, L.~Beyer, A.~Kolesnikov, D.~Weissenborn, X.~Zhai,
  T.~Unterthiner, M.~Dehghani, M.~Minderer, G.~Heigold, S.~Gelly \emph{et~al.},
  ``An image is worth 16x16 words: Transformers for image recognition at
  scale,'' \emph{arXiv preprint arXiv:2010.11929}, 2020.

\bibitem{xiao2023baseg}
X.~Xiao, Y.~Zhao, F.~Zhang, B.~Luo, L.~Yu, B.~Chen, and C.~Yang, ``Baseg:
  Boundary aware semantic segmentation for autonomous driving,'' \emph{Neural
  Networks}, vol. 157, pp. 460--470, 2023.

\bibitem{chen2022utrad}
L.~Chen, Z.~You, N.~Zhang, J.~Xi, and X.~Le, ``Utrad: Anomaly detection and
  localization with u-transformer,'' \emph{Neural Networks}, vol. 147, pp.
  53--62, 2022.

\bibitem{xu2021dc}
R.~Xu, C.~Wang, S.~Xu, W.~Meng, and X.~Zhang, ``Dc-net: Dual context network
  for 2d medical image segmentation,'' in \emph{International Conference on
  Medical Image Computing and Computer-Assisted Intervention}.\hskip 1em plus
  0.5em minus 0.4em\relax Springer, 2021, pp. 503--513.

\bibitem{wang2022net}
C.~Wang, R.~Xu, S.~Xu, W.~Meng, and X.~Zhang, ``Da-net: Dual branch transformer
  and adaptive strip upsampling for retinal vessels segmentation,'' in
  \emph{International Conference on Medical Image Computing and
  Computer-Assisted Intervention}.\hskip 1em plus 0.5em minus 0.4em\relax
  Springer, 2022, pp. 528--538.

\bibitem{valanarasu2021medical}
J.~M.~J. Valanarasu, P.~Oza, I.~Hacihaliloglu, and V.~M. Patel, ``Medical
  transformer: Gated axial-attention for medical image segmentation,'' in
  \emph{International Conference on Medical Image Computing and
  Computer-Assisted Intervention}.\hskip 1em plus 0.5em minus 0.4em\relax
  Springer, 2021, pp. 36--46.

\bibitem{chen2021transunet}
J.~Chen, Y.~Lu, Q.~Yu, X.~Luo, E.~Adeli, Y.~Wang, L.~Lu, A.~L. Yuille, and
  Y.~Zhou, ``Transunet: Transformers make strong encoders for medical image
  segmentation,'' \emph{arXiv preprint arXiv:2102.04306}, 2021.

\bibitem{wang2021boundary}
J.~Wang, L.~Wei, L.~Wang, Q.~Zhou, L.~Zhu, and J.~Qin, ``Boundary-aware
  transformers for skin lesion segmentation,'' in \emph{International
  Conference on Medical Image Computing and Computer-Assisted
  Intervention}.\hskip 1em plus 0.5em minus 0.4em\relax Springer, 2021, pp.
  206--216.

\bibitem{zhang2018context}
H.~Zhang, K.~Dana, J.~Shi, Z.~Zhang, X.~Wang, A.~Tyagi, and A.~Agrawal,
  ``Context encoding for semantic segmentation,'' in \emph{Proceedings of the
  IEEE conference on Computer Vision and Pattern Recognition}, 2018, pp.
  7151--7160.

\bibitem{zhang2017deep}
H.~Zhang, J.~Xue, and K.~Dana, ``Deep ten: Texture encoding network,'' in
  \emph{Proceedings of the IEEE conference on computer vision and pattern
  recognition}, 2017, pp. 708--717.

\bibitem{wang2016learnable}
Z.~Wang, H.~Li, W.~Ouyang, and X.~Wang, ``Learnable histogram: Statistical
  context features for deep neural networks,'' in \emph{European Conference on
  Computer Vision}.\hskip 1em plus 0.5em minus 0.4em\relax Springer, 2016, pp.
  246--262.

\bibitem{xie2015effective}
J.~Xie, L.~Zhang, J.~You, and S.~Shiu, ``Effective texture classification by
  texton encoding induced statistical features,'' \emph{Pattern recognition},
  vol.~48, no.~2, pp. 447--457, 2015.

\bibitem{zhu2021learning}
L.~Zhu, D.~Ji, S.~Zhu, W.~Gan, W.~Wu, and J.~Yan, ``Learning statistical
  texture for semantic segmentation,'' in \emph{Proceedings of the IEEE/CVF
  Conference on Computer Vision and Pattern Recognition}, 2021, pp.
  12\,537--12\,546.

\bibitem{sinha2020multi}
A.~Sinha and J.~Dolz, ``Multi-scale self-guided attention for medical image
  segmentation,'' \emph{IEEE journal of biomedical and health informatics},
  vol.~25, no.~1, pp. 121--130, 2020.

\bibitem{ruan2022malunet}
J.~Ruan, S.~Xiang, M.~Xie, T.~Liu, and Y.~Fu, ``Malunet: A multi-attention and
  light-weight unet for skin lesion segmentation,'' in \emph{2022 IEEE
  International Conference on Bioinformatics and Biomedicine (BIBM)}.\hskip 1em
  plus 0.5em minus 0.4em\relax IEEE, 2022, pp. 1150--1156.

\bibitem{woo2018cbam}
S.~Woo, J.~Park, J.-Y. Lee, and I.~So~Kweon, ``Cbam: Convolutional block
  attention module,'' in \emph{Proceedings of the European conference on
  computer vision (ECCV)}, 2018, pp. 3--19.

\bibitem{milletari2016loss}
F.~Milletari, N.~Navab, and S.-A. Ahmadi, ``V-net: Fully convolutional neural
  networks for volumetric medical image segmentation,'' in \emph{2016 fourth
  international conference on 3D vision (3DV)}.\hskip 1em plus 0.5em minus
  0.4em\relax IEEE, 2016, pp. 565--571.

\bibitem{wang2021pr2}
X.~Wang, X.~Jiang, H.~Ding, Y.~Zhao, and J.~Liu, ``Knowledge-aware deep
  framework for collaborative skin lesion segmentation and melanoma
  recognition,'' \emph{Pattern Recognition}, vol. 120, p. 108075, 2021.

\bibitem{gu2020canet}
R.~Gu, G.~Wang, T.~Song, R.~Huang, M.~Aertsen, J.~Deprest, S.~Ourselin,
  T.~Vercauteren, and S.~Zhang, ``Ca-net: Comprehensive attention convolutional
  neural networks for explainable medical image segmentation,'' \emph{IEEE
  transactions on medical imaging}, vol.~40, no.~2, pp. 699--711, 2020.

\bibitem{dai2022ms}
D.~Dai, C.~Dong, S.~Xu, Q.~Yan, Z.~Li, C.~Zhang, and N.~Luo, ``Ms red: A novel
  multi-scale residual encoding and decoding network for skin lesion
  segmentation,'' \emph{Medical Image Analysis}, vol.~75, p. 102293, 2022.

\bibitem{valanarasu2021medt}
J.~M.~J. Valanarasu, P.~Oza, I.~Hacihaliloglu, and V.~M. Patel, ``Medical
  transformer: Gated axial-attention for medical image segmentation,'' in
  \emph{International Conference on Medical Image Computing and
  Computer-Assisted Intervention}.\hskip 1em plus 0.5em minus 0.4em\relax
  Springer, 2021, pp. 36--46.

\bibitem{wang2021bat}
J.~Wang, L.~Wei, L.~Wang, Q.~Zhou, L.~Zhu, and J.~Qin, ``Boundary-aware
  transformers for skin lesion segmentation,'' in \emph{International
  Conference on Medical Image Computing and Computer-Assisted
  Intervention}.\hskip 1em plus 0.5em minus 0.4em\relax Springer, 2021, pp.
  206--216.

\bibitem{aghdam2023attention}
E.~K. Aghdam, R.~Azad, M.~Zarvani, and D.~Merhof, ``Attention swin u-net:
  Cross-contextual attention mechanism for skin lesion segmentation,'' in
  \emph{2023 IEEE 20th International Symposium on Biomedical Imaging
  (ISBI)}.\hskip 1em plus 0.5em minus 0.4em\relax IEEE, 2023, pp. 1--5.

\bibitem{yuan2017improving}
Y.~Yuan and Y.-C. Lo, ``Improving dermoscopic image segmentation with enhanced
  convolutional-deconvolutional networks,'' \emph{IEEE journal of biomedical
  and health informatics}, vol.~23, no.~2, pp. 519--526, 2017.

\bibitem{isic2017}
ISIC, ``Skin lesion analysis towards melanoma detection,''
  \emph{https://challenge.kitware.com/\#phase/584b0afacad3a51cc66c8e24}, 2017.

\bibitem{li2018dense}
H.~Li, X.~He, F.~Zhou, Z.~Yu, D.~Ni, S.~Chen, T.~Wang, and B.~Lei, ``Dense
  deconvolutional network for skin lesion segmentation,'' \emph{IEEE journal of
  biomedical and health informatics}, vol.~23, no.~2, pp. 527--537, 2018.

\bibitem{wang2019bi}
X.~Wang, X.~Jiang, H.~Ding, and J.~Liu, ``Bi-directional dermoscopic feature
  learning and multi-scale consistent decision fusion for skin lesion
  segmentation,'' \emph{IEEE transactions on image processing}, vol.~29, pp.
  3039--3051, 2019.

\bibitem{xie2020mutual}
Y.~Xie, J.~Zhang, Y.~Xia, and C.~Shen, ``A mutual bootstrapping model for
  automated skin lesion segmentation and classification,'' \emph{IEEE
  transactions on medical imaging}, vol.~39, no.~7, pp. 2482--2493, 2020.

\bibitem{ji2021multi}
Y.~Ji, R.~Zhang, H.~Wang, Z.~Li, L.~Wu, S.~Zhang, and P.~Luo, ``Multi-compound
  transformer for accurate biomedical image segmentation,'' in
  \emph{International Conference on Medical Image Computing and
  Computer-Assisted Intervention}.\hskip 1em plus 0.5em minus 0.4em\relax
  Springer, 2021, pp. 326--336.

\bibitem{tschandl2018ham10000}
P.~Tschandl, C.~Rosendahl, and H.~Kittler, ``The ham10000 dataset, a large
  collection of multi-source dermatoscopic images of common pigmented skin
  lesions,'' \emph{Scientific data}, vol.~5, no.~1, pp. 1--9, 2018.

\end{thebibliography}

\end{document}